\documentclass[utf8]{frontiersSCNS} % for Science, Engineering and Humanities and Social Sciences articles

\setcitestyle{square} % for Physics and Applied Mathematics and Statistics articles
\usepackage{url,hyperref,lineno,microtype,subcaption}
\usepackage[onehalfspacing]{setspace}
\usepackage{siunitx}
\usepackage{xspace}

\newcommand{\NEW}{NEXT-White}
\newcommand{\NEXT}{NEXT-100}

\newcommand{\fig}{figure}

\newcommand{\eg}{{\it e.g.}}

\newcommand{\micro}{\ensuremath{\mu}}

\newcommand{\bb}{\ensuremath{\beta\beta}}

\newcommand{\bbonu}{\ensuremath{\beta\beta0\nu}}
\newcommand{\bbtnu}{\ensuremath{\beta\beta2\nu}}
\newcommand{\tz}{\ensuremath{t_0}}

\newcommand{\mbb}{\ensuremath{m_{\beta\beta}}}

\newcommand{\ckky}{\ensuremath{\rm counts/(keV \cdot kg \cdot y)}}

\newcommand{\Qbb}{\ensuremath{Q_{\beta\beta}}}

\newcommand{\Tonu}{\ensuremath{T_{1/2}^{0\nu}}}

\newcommand{\CS}{\ensuremath{^{137}}\rm Cs}

\newcommand{\NA}{\ensuremath{^{22}}\rm Na}

\newcommand{\PBD}{\ensuremath{^{210}}\rm Pb}

\newcommand{\PO}{\ensuremath{^{214}}\rm Po}

\newcommand{\RAD}{\ensuremath{^{222}}\rm Rn}

\newcommand{\XE}{\ensuremath{{}^{136}\rm Xe}}
\newcommand{\GE}{\ensuremath{{}^{76}\rm Ge}}

\newcommand{\TL}{\ensuremath{{}^{208}\rm{Tl}}}

\newcommand{\COFiftySeven}{\ensuremath{{}^{57}\rm Co}}

\newcommand{\BI}{\ensuremath{{}^{214}}{\rm Bi}}
\newcommand{\Bapp}{\ensuremath{{\rm Ba}^{++}}}
\newcommand{\Bap}{\ensuremath{{\rm Ba}^{+}}}

\newcommand{\Capp}{\ensuremath{{\rm Ca}^{++}}}

\newcommand{\K}[1]{\ensuremath{^{#1}\mathrm{K}}\xspace}

\DeclareSIUnit\c{\mbox{$c$}}
\DeclareSIUnit\magn{\mbox{$\times$}}
\DeclareSIUnit\min{min}
\DeclareSIUnit\week{week}
\DeclareSIUnit\year{yr}
\DeclareSIUnit\years{years}
\DeclareSIUnit\yr{yr}
\DeclareSIUnit\standard{std}
\DeclareSIUnit\str{sr}
\DeclareSIUnit\ppm{ppm}
\DeclareSIUnit\ppb{ppb}
\DeclareSIUnit\ppt{ppt}
\DeclareSIUnit\pe{PE}
\DeclareSIUnit\spe{SPE}
\DeclareSIUnit\ev{events}
\DeclareSIUnit\ct{counts}
\DeclareSIUnit\neutron{\mbox{$n$}}
\DeclareSIUnit\smp{samples}
\DeclareSIUnit\Sample{S}
\DeclareSIUnit\ch{ch}
\DeclareSIUnit\hit{hit}
\DeclareSIUnit\hits{hits}
\DeclareSIUnit\bin{(\mbox{5-PE}~bin)}
\DeclareSIUnit\sgm{\mbox{$\sigma$}}
\DeclareSIUnit\rms{RMS}
\DeclareSIUnit\keVr{\mbox{keV$_{\rm nr}$}}
\DeclareSIUnit\keVee{\mbox{keV$_{e{\rm e}}$}}
\DeclareSIUnit\ph{photon}
\DeclareSIUnit\pes{pes}
\DeclareSIUnit\el{electrons}
\DeclareSIUnit\pm{PMT}
\DeclareSIUnit\inch{"}
\DeclareSIUnit\bit{bit}
\DeclareSIUnit\sample{samples}
\DeclareSIUnit\barn{barn}
\DeclareSIUnit\bara{bar}
\DeclareSIUnit\barg{barg}
\DeclareSIUnit\mlardepth{\mbox(meter~of~\LAr~depth)}
\DeclareSIUnit\Curie{Ci}
\DeclareSIUnit\psi{psi}
\DeclareSIUnit\parsec{pc}
\DeclareSIUnit\liveday{\mbox{live-days}}
\DeclareSIUnit\days{\mbox{days}}
\DeclareSIUnit\day{\mbox{day}}
\DeclareSIUnit\miles{\mbox{miles}}
\DeclareSIUnit\degreeC{\mbox{$^{\circ}$C}}
\DeclareSIUnit\electron{\mbox{$e^-$}}
\DeclareSIUnit\Euro{\mbox{\euro}}
\DeclareSIUnit\cph{cph}
\DeclareSIUnit\neq{neq}
\DeclareSIUnit\unit{unit}
\DeclareSIUnit\byte{Byte}
\DeclareSIUnit\Bq{\becquerel}

\newcommand{\XeWaveLength}{\SI{172}{\nano\meter}}

\newcommand{\BackgroundFreeLimit}{\SI{0.1}{\ev}}
\newcommand{\AlmostBackgroundFreeLimit}{\SI{<1}{\ev}}

\newcommand{\Next}{\mbox{NEXT-100}}

\newcommand{\Ntk}{\mbox{NEXT-2.0}}

\newcommand{\HPXeEL}{HPXe-EL}

\newcommand{\TPB}{Tetraphenyl Butadiene}
\newcommand{\PDOT}{Poly-Ethylenedioxythiophene}

\newcommand{\XenonDensity}{\SI{5.761}{\kg\per\cubic\meter}}

\newcommand{\XenonEnergyPerElectron}{\SI{21.9}{\eV}}

\newcommand{\XenonQbb}{\SI{2458}{\keV}}

\newcommand{\MCDriftVelocity}{\SI{1}{\mm\per\micro\second}}

\newcommand{\KrEnergy}{\SI{41.5}{\keV}}

\newcommand{\NextTrackLengthAtQbb}{\SI{15}{\cm}}

\newcommand{\TransverseDiffusionPureXenon}{\SI{10}{\mm\per\sqrt\m}}
\newcommand{\TransverseDiffusionXeHe}{\SI{2}{\mm\per\sqrt\m}}

\newcommand{\NewPressureVesselMaterial}{316Ti}

\newcommand{\NewTpcLength}{\SI{664.5}{\mm}}

\newcommand{\NewTpcDriftLength}{\SI{530.3 +- 2}{\mm}}

\newcommand{\NewFieldCageHDPEThickness}{\SI{21}{\mm}}

\newcommand{\NewDriftField}{\SI{400}{\V\per\cm}}

\newcommand{\NewTpcELGap}{\SI{6}{\mm}}
\newcommand{\NewAnodePlateDiameter}{\SI{522}{\mm}}
\newcommand{\NewAnodePlateThickness}{\SI{3}{\mm}}

\newcommand{\NewNumberOfSiPM}{\num{1792}}

\newcommand{\NewNumberOfBoards}{\num{28}}
\newcommand{\NewNumberOfSiPMPerBoard}{8 x 8}
\newcommand{\NewSiPMSeries}{SensL C}
\newcommand{\NewSiPMModel}{MicroFC-10035-SMT-GP}
\newcommand{\NewSiPMSize}{\SI{1}{\mm\square}}
\newcommand{\NewSipmPitch}{\SI{10}{\mm}}
\newcommand{\NewSipmCell}{\SI{35}{\micro\meter}}
\newcommand{\NewSipmDarkCount}{\SI{100}{\kHz}}
\newcommand{\NewPhotoelectronsPerSiPM}{\SI{250}{\pes\per\micro\second}}
\newcommand{\TrackingPlaneToEL}{\SI{8}{\mm}}
\newcommand{\TrackingPlaneToAnode}{\SI{2}{\mm}}
\newcommand{\NewTrackingPlaneEndCapThickness}{\SI{120}{\mm}}

\newcommand{\NewNumberOfPMT}{\num{12}}

\newcommand{\NewTpcDiameter}{\SI{454}{\mm}}

\newcommand{\NewPressure}{\SI{15}{\bar}}

\newcommand{\NewPmtEndCapThickness}{\SI{120}{\mm}}

\newcommand{\NewTypePMT}{Hamamatsu R11410-10}

\newcommand{\NewPMTActivity}{\SI{0.37 +- 0.08}{\milli\becquerel\per\unit}}

\newcommand{\NewK}{0.016}

\newcommand{\XeEnrichment}{\SI{90}{\percent}}

\newcommand{\NextTpcDiameter}{\SI{1050}{\mm}}
\newcommand{\NextTpcLength}{\SI{1300}{\mm}}
\newcommand{\NextFiducialVolume}{\SI{1.27}{\cubic\meter}}

\newcommand{\NextFiducialMass}{\SI{97}{\kg}}

\newcommand{\NextPressure}{\SI{15}{\bar}}
\newcommand{\NextTemperature}{\SI{293}{\K}}

\newcommand{\NextNumberOfSiPM}{\num{5600}}
\newcommand{\NextNumberOfPMT}{\num{60}}

\newcommand{\NtkOperatingTemperature}{\SI{175}{\K}}

\newcommand{\NtkTimesNext}{5}

\def\keyFont{\fontsize{8}{11}\helveticabold }
\def\firstAuthorLast{J.J. Gomez-Cadenas {et~al.}} %use et al only if is more than 1 author
\def\Authors{J.J. Gomez-Cadenas\,$^{1,2,*}$, F. Monrabal\,$^{1}$ and P. Ferrario\,$^{1,2,*}$}
\def\Address{$^{1}$Donostia International Physics Center (DIPC), Spain. \\
$^{2}$Ikerbasque, Spain.  }
\def\corrAuthor{Corresponding Author}

\def\corrEmail{jjgomezcadenas@dipc.org}

\begin{document}
\onecolumn
\firstpage{1}

\title[High Pressure Gas Xenon TPCs for Double Beta Decay Searches]{High Pressure Gas Xenon TPCs for Double Beta Decay Searches} 

\author[\firstAuthorLast ]{\Authors} %This field will be automatically populated
\address{} %This field will be automatically populated
\correspondance{} %This field will be automatically populated

\extraAuth{}% If there are more than 1 corresponding author, comment this line and uncomment the next one.
%\extraAuth{corresponding Author2 \\ Laboratory X2, Institute X2, Department X2, Organization X2, Street X2, City X2 , State XX2 (only USA, Canada and Australia), Zip Code2, X2 Country X2, email2@uni2.edu}

\maketitle

\begin{abstract}

\section{}
This article reviews the application of high pressure gas xenon (HPXe) time projection chambers to neutrinoless double beta decay experiments. First, the fundamentals of the technology and the historical development of the field are discussed. Then, the state of the art is presented, including the prospects for the next generation of experiments with masses in the ton scale. 

\tiny
 \keyFont{ \section{Keywords:} xenon, neutrinos, electroluminescence, resolution, topology, barium tagging, SMFI.} 
 %All article types: you may provide up to 8 keywords; at least 5 are mandatory.
\end{abstract}

\section{Introduction}

The invention of the time projection chamber (TPC) \citep{Nygren:1976fe} revolutionized the imaging of charged particles in gaseous detectors. More than four decades after its introduction, the TPC is still one of the most important detectors in particle physics.
% and a fundamental revolution in rare event searches.
 
Over the last decade, xenon TPCs have emerged as powerful tools for the study of rare events, in particular concerning dark matter and 
neutrinoless double decay (\bbonu) searches. Their principle of operation is the same as for all TPCs. Charged radiation ionizes the fluid and the ionization electrons are drifted under the action of an electric field to sensitive image planes, where their transverse position information X,Y is collected. Their arrival times (relative to the start-of-the-event time, or \tz) are then traded to longitudinal positions, Z, through their average drift velocity. Yet, the application of TPCs to \bbonu\ searches has its own peculiarities. In this case xenon is not only the sensitive medium, but also the target where the decays occur. Since the sensitivity of the search is proportional to the target mass the apparatus needs to be as large and compact as possible, leading to either high pressure xenon (HPXe) or liquid xenon (LXe) TPCs. Furthermore, the energy regime is relatively low (the end-point of the decay 
Xe $\rightarrow \Bapp + 2 e^-$, \Qbb, is  \XenonQbb) and thus the tracks left by the two electrons can be rather short for HPXe detectors (of the order of \NextTrackLengthAtQbb\ for electrons
with \Qbb\ energies at \NextPressure) or even point-like objects for LXe chambers. Both types of TPCs act as calorimeters, capable of measuring the total energy of the decay and to locate the interaction in a well defined fiducial volume thanks to the availability of a mechanism to signal \tz, namely the VUV scintillation emitted by xenon as a response to ionizing radiation. In addition a HPXe TPC provides a {\em topological signature}, thanks to its capability to image the electron tracks. 
 
In this review we discuss the fundamentals, state-of-the-art and potential for the next generation of experiments searching for neutrinoless double beta decay processes with high pressure xenon TPCs.  A more general discussion encompassing the various TPCs used for rare event searches can be found in \citep{Gonzalez-Diaz:2017gxo}.
 
Neutrinoless double beta decay is a hypothetical, very slow radioactive process in which two neutrons undergo $\beta$-decay simultaneously and without the emission of neutrinos, 
$(Z,A) \rightarrow (Z+2,A) + 2\ e^{-}$.  An unambiguous observation of this process would establish that neutrinos are Majorana particles \citep{Majorana:1937vz}, identical to their antiparticles and would have deep implications in physics and cosmology~\citep{Fukugita:1986bw}.
 
 The simplest mechanism to mediate such a transition is the virtual exchange of light Majorana neutrinos. Assuming this exchange to be  dominant at low energies, the half-life of \bbonu\ can be written as:
\begin{equation}
(\Tonu)^{-1} = G^{0\nu} \ \big|M^{0\nu}\big|^{2} \ \mbb^{2}; \,\,\, \,\,
\label{eq:to_mbb}
\end{equation}
where $G^{0\nu}$ is a phase-space integral for the emission of two electrons, $M^{0\nu}$~ is the nuclear matrix element (NME) of the transition, and \mbb\ is the \emph{effective Majorana mass} of the electron neutrino, defined in terms of the neutrino mass eigenstates ($m_{i}$) and the elements of the neutrino mixing matrix \citep{GomezCadenas:2012fe}.

Over the last decade, several 
\bbonu\ experiments, with masses in the range of few tens to few hundreds of kilograms have pushed the sensitivity to the half-life of \bbonu\ processes by more than one order of magnitude in three different isotopes.  Four of these experiments (GERDA~\citep{Agostini:2018tnm}, EXO~\citep{Albert:2017owj}, KamLAND-Zen~\citep{Gando:2016br, Gando:2016ji}, and CUORE~\citep{Alduino:2017wj}) have recently published the results of their analysis.  The best limit on the lifetime of a \bbonu\ isotope was obtained by KamLAND-Zen~\citep{Gando:2016br, Gando:2016ji}), corresponding to $\Tonu > \SI{1.07E26}{\yr}$ for the \bbonu\ decay of \XE. The GERDA experiment has recently published a limit $\Tonu > \SI{0.8E26}{\yr}$ for the \bbonu\ decay of \GE.

The target of the ``next generation'' of xenon-experiment is to improve the sensitivity to \Tonu\ by roughly one order of magnitude, to some \SI{E27}{\yr}, or about 
\SI{~20}{\meV} in \mbb. If the neutrino mass hierarchy is inverted (\eg\ $\Delta m_{31}^2 <0$, where $\Delta m_{ij}^2 = m_i^2 - m_j^2$~and $m_i, i=1,3$~denotes the mass of the three neutrino mass eigenstates), reaching such a sensitivity on \mbb\ would result in a discovery if the neutrino is a Majorana particle. Even if Nature has chosen a normal hierarchy ($\Delta m_{31}^2 >0$) a statistical analysis ~\citep{Agostini:2017jim} suggests that the probability of discovery would be rather large, around 50\%. 

Reaching a sensitivity of  \SI{E27}{\yr} in the half-life \Tonu\
will require larger exposures (the product of fiducial mass $M$ and observation time $t$), and thus larger detectors than those of the current generation. The number of events
observed in a detector containing a mass $M$~of a \bbonu\ decaying isotope of atomic weight $W$~and taking data over a period of time $t$ is related with \Tonu\ through:

\begin{equation}
\Tonu = \epsilon \log{2} \frac{N_A M t}{W N_{\beta\beta} }
\label{eq:to}
\end{equation}
where $N_A$~is the Avogadro number and $\epsilon$ the detector efficiency. In the absence of background, the observation of a single event would determine the existence of the process and measure the value of \Tonu.  
For example, if $\Tonu = 10^{27}$~year, the observation of 1 event in one year in a detector with $30\%$ efficiency would require a mass of 1 tonne.
%If $\Tonu = 10^{27}$~year, the observation of 1 event per year would require an exposure of 300 kg, for example a detector with a mass of \SI{1}{ton} of xenon and an overall detector efficiency of 30\%.
%, one event per year would be observed if $\Tonu = 10^{27}$~year.

In the presence of backgrounds, however, the sensitivity to \Tonu\ will scale like
$1/\sqrt{N}$~(where N is the number of observed events) rather than scaling like $1/N$~as in the case of a background-free experiment. Consequently, the sensitivity to \Tonu\ improves with $\sqrt{Mt}$~rather than with $Mt$. Alas, equation \ref{eq:to_mbb} dictates that the sensitivity to the physical parameter (the effective neutrino mass \mbb) goes with
the square root of the half-life, and thus each order of magnitude improvement in the latter brings in only a factor 3 improvement in the former. In an experiment where backgrounds need to be subtracted, on the other hand, one needs two orders of magnitude increase in the exposure $M t$ to improve one order of magnitude in \Tonu\ . 
% which buys a mere factor 3 in \mbb.  %%It was repeated
It follows that {\em the next generation of \bbonu\ experiments must feature target masses in the tonne-range, while aiming to reduce backgrounds to virtually zero}.

Among all the \bb\ decay isotopes where \bbonu\ processes could occur, \XE\ is the cheapest and easiest to obtain. 
%This can be simply established by the fact that the total amount of enriched xenon (90\% of \XE) in the world exceeds already one ton (\SI{800}{kg}  owned by the KamLAND-Zen collaboration in Japan, \SI{200}{kg} owned by the EXO collaboration in the USA and \SI{100}{kg} owned by the NEXT experiment in Spain\footnote{In fact the xenon is property of the Canfranc Underground Laboratory, LSC, and borrowed by the NEXT collaboration}), while there is in the world less than \SI{50}{kg} of enriched germanium (90\% of \GE), an isotope which has been the battle horse of \bbonu\ searches for decades. 
%
Furthermore, xenon is a noble gas that can be dissolved in liquid scintillator (the approach of KamLAND-Zen) or used to build TPCs. The EXO collaboration has pioneered the LXe technology, while the NEXT program \citep{Cadenas:2017vcu} is leading the development of the HPXe technology. 
%All the xenon-based detectors enjoy economy of scale (\eg\, detectors with roughly 10 times more mass can be built by doubling the longitudinal dimensions of the apparatus) and a ratio signal/background (S/N) which tends to improve as the detector grows (since the target mass goes with the volume while the backgrounds tend to concentrate in surfaces). 

When compared with the other xenon-based experiments, the HPXe technology has the advantage of very good intrinsic energy resolution 
%(when using electroluminescence to amplify the ionization signal) I prefeer to remove this as people can wonder why EXO does not have good Eres
, and the availability of a topological signature (the observation of the two electrons characteristic of the \bbonu\ decay) that permits a very low background count in the region of interest (ROI) near \Qbb. The main disadvantages is a relatively lower selection efficiency, of the order of $30\%$, mostly due to the losses of events that radiate bremsstrahlung photons and to the cost of imposing topological recognition. Although not as compact as LXe TPCs, at sufficiently large pressures (\eg \NextPressure) a HPXe of relatively modest size (about \SI{10}{\cubic\meter} of volume) can host masses in the ton scale.
Furthermore, the possibility of tagging the \Bapp\ nuclei produced in the \bbonu\ decay of xenon opens up the possibility of background free searches for xenon-based TPCs. 

This review is organized as follows. Fundamentals are discussed in section \ref{sec.fundamentals}. A quick historical review of the field is presented
in \ref{sec.history}. Section \ref{sec.next} is devoted to the NEXT program. In section \ref{sec.axel} the AXEL and PANDA-X-III proposals are described. Section \ref{sec.smfi} describes the on-going efforts in barium tagging in HPXe detectors. An outlook is
presented in section \ref{sec.conclu}.

\section{Fundamentals}
\label{sec.fundamentals}

\subsection{Operational parameters of a HPXe TPC: pressure, temperature and density}
\label{sec.operation}

%\begin{figure}[hb!]
%\centering
%\includegraphics[width=0.5\textwidth]{img/p_vs_t_xe.png}
%\caption{\small Isochoric curves for xenon at different densities.}
%\label{fig.isxe}
%\end{figure}

So far, HPXe detectors have operated at ambient temperature with pressures varying between \SI{5}{\bar} ---St.Gotthard TPC \citep{Iqbal:1987vh}--- and \SI{20}{\bar} ---NEXT-DBDM prototype \citep{Alvarez:2012hh}---. The \NEW\ detector \citep{Monrabal:2018xlr} is currently taking data at \SI{10}{\bar}. In practice, at standard temperature, the operational pressure for ton-scale detectors will be in the range \SIrange{10}{20}{\bar}.  Given the density of xenon gas at a pressure of \SI{1}{bar} and \SI{300}{K} temperature (\XenonDensity), a HPXe TPC of  \SI{10}{\cubic\meter} operating at \SI{10}{\bar} would hold a target mass of near \SI{600}{kg}
( \SI{1.2}{ton} at  \SI{20}{\bar}). The detector dimensions, while large (\eg\ \SI{3.2}{\meter} length by \SI{3}{\meter} diameter) appear as technically feasible. On the other hand, the
NEXT demonstrators (NEXT-DEMO, NEXT-DBDM and \NEW) have shown excellent energy resolution and a powerful topological signature in this pressure range.

\begin{figure}[h!]
\centering
\includegraphics[angle=-90, width=1.0\textwidth]{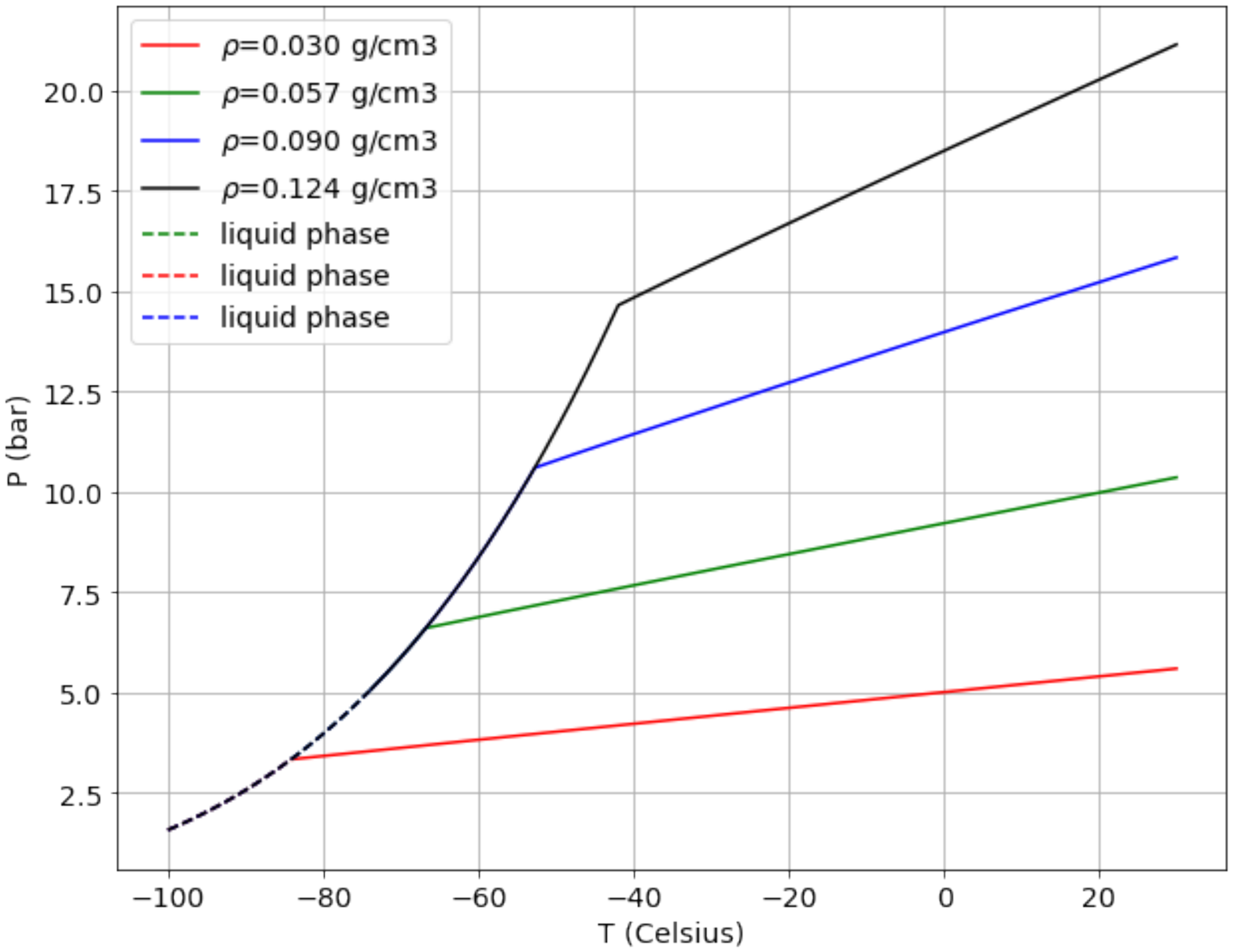}
\caption{\small Isochoric curves for xenon at different densities.}
\label{fig.isxe}
\end{figure}

However, pressure can be traded with temperature, as illustrated in \fig\ \ref{fig.isxe}, which shows the four isochoric (constant density) curves corresponding to pressures of $5$, $10$, $15$ and $20$ bar (at a temperature of \SI{20}{\celsius}). An interesting possibility would be cooling the detector to temperatures near the liquefaction point (\eg\ \SI{-70}{\celsius} for a density of
\SI{0.057}{\gram\per\cubic\cm}). The advantages of operating at low temperatures will be discussed in section \ref{sec.next}.

\subsection{Ionization}
\label{sec.signal}

When a charged particle propagates through a noble gas, the Coulomb interaction with the atoms results in {\it ionization} of the medium, releasing on average $\bar{n}_e$
electron-ion pairs, and $N_{ex}$~excited atoms. Sub-excitation electrons, (\eg\ free electrons with a kinetic energy lower than the energy of the first excited level), are also produced. This can be expressed as~\citep{Aprile:2009dv}:
\begin{equation}
E =N_i E_i + N_{ex} E_{ex} + N_i \epsilon,
\end{equation}
where $E$~is the energy deposited in the medium in the form of ionization, excitation, and sub-excitation electrons; $N_i$ ~ is the number of electron-ion pairs produced at an average energy deposition $E_i$, $N_{ex}$~ is the number of excited atoms produced at an average energy deposition $E_{ex}$ and $\epsilon$~ is the average kinetic energy of sub-excitation electrons. Then the expected number of electrons produced for an energy deposition $E$~ is:

%The average energy required to produce one electron-ion pair, $W_I$ is defined as:
%%
%\begin{equation}
%W_I = E_0/N_i = E_i + E_{ex} (N_{ex}/N_i) + \epsilon
%\end{equation}
%%
%and thus, the average number of electron-ion pairs produced when a particle releases its
%energy $E$~in the gas is:

\begin{equation}
\bar{n}_e =\frac{E}{W_I}
\label{eq:ne}
\end{equation}
where $W_I$~ is the average energy required to produce one electron-ion pair.
In xenon gas $W_I = \XenonEnergyPerElectron$~\citep{Aprile:2009dv}. A \bbonu\ event of energy \XenonQbb\ results, therefore, in the average production of $112\ 237$ electron-ion pairs.
In a HPXe TPC a moderate electric field will drift the electrons towards the anode and the ions towards the cathode, minimizing recombination.

%An approximate formulas to express the effect of recombination is discussed in  \citep{Gonzalez-Diaz:2017gxo}:
%
%\begin{equation}
%(1-\mathcal{R}) \sim \frac{E_d}{k'}\ln(1+\frac{k'}{E_d}) \label{BoxModel}
%\end{equation}
%where $k'$ is a fitting constant, and $E_d$~is the applied electric field.
%
%The $W_I$ value obtained for pure gases under low ionizing particles, e.g., ``minimum ionizing particles'' (mips), keV-electrons and x-rays is a rather well defined magnitude, irrespective from the particle energy and the gas pressure \citep{ICRU}. Additionally, and as charge recombination is usually very small in those cases (\citep{Bolot},\citep{Balan}). The electrons emitted in \bbonu\ decays have energies in the range of thousand keV (given the value $\Qbb = \XenonQbb$. At these energies, electrons can loose energy by Bremsstrahlung and emission of energetic $\delta$ rays, in addition to ionization. Bremsstrahlung emission causes a significant efficiency loss for HPXe detectors, since a number of \bbonu\ events will radiate an energetic photon ---which often escapes detection--- resulting in a measured energy outside region of interest (ROI), defined by the detector resolution.

%It must be noted that $W_I$~ accounts for the ionization produced directly by the impinging particle, as well as that produced by the released electrons themselves. This results in the creation of ionization `clusters', that broaden the spatial distribution of the ionization trail. It also means that both direct and secondary ionization processes are present in general.

\subsection{Scintillation}
The propagation of a charged particle in a noble gas also results in the emission of VUV scintillation light (with an average wavelength of \XeWaveLength\ in xenon). Defining $W_s$ as the average energy needed in the creation of one primary scintillation photon, the average number of scintillation photons produced when a particle releases its
energy $E$~in the gas is:

\begin{equation}
\bar{n}_\gamma =\frac{E}{W_s}
\label{eq:ng}
\end{equation}
The NEXT collaboration has measured the value of $W_s$ in xenon gas to be \citep{Fernandes:2010gg}:

\begin{equation}
	W_s = 76 \pm 6 ~\mathrm{eV}
\label{eq:Ws}
\end{equation}

Thus a \bbonu\ event will release on average $32\ 342$ photons.
Since light production is isotropic, only a small fraction of the produced photons, $\Omega$, (typically of the order of few \%) can be collected. Measurement of the primary scintillation, however, is crucial for a \bbonu\ detector, as it signals \tz. A measurement of \tz\ is essential to fiducialize the events and remove the large rate of background events that accumulate at the electrodes, and to correct for charge losses occurring during charge drift. Without such corrections, the performance of the detector both in terms of background rate and resolution is seriously compromised.

%The need to record \tz\ has two major implications. The first one is that an HPXe has to be instrumented with photosensors covering an area large enough for the collection of the scintillation signal. The second implication is that only mixtures that do not quench the scintillation light ---or mixtures capable to absorb and re-emit the scintillation light without introducing additional fluctuations--- are acceptable. As we will see later, the design of the NEXT detectors (as well as the AXEL demonstrator) follow this two constrains, while the St. Gottard TPC did not, paying a dearly price for it in terms of background and energy resolution, a fate that appears also unavoidable for the PANDAS-X-III proposal, as discussed in section XXX.

\subsection{Diffusion}
As the ionization electrons (and the positive ions) drift towards the anode (cathode) under the action of the electric field, they interact with the noble gas atoms, resulting in both longitudinal and transverse diffusion.
%Both coefficients appear in the equation describing the propagation of the ionization charge, obtained through the hydrodynamic approximation of the Boltzmann equations \cite{Huxley}:

%\begin{eqnarray}
%&\frac{\partial \mathcal{N}_e}{\partial t} &+ v_d \frac{\partial \mathcal{N}_e}{\partial z'} - D_T\bigg(\frac{\partial^2 \mathcal{N}_e}{\partial x'^2} + \frac{\partial^2 \mathcal{N}_e}{\partial y'^2}\bigg) - D_L\frac{\partial^2 \mathcal{N}_e}{\partial z'^2} \nonumber \\
%&&= -\eta v_d \mathcal{N}_e \label{Eq_hydro}
%\end{eqnarray}
Defining
$\mathcal{N}_e$~ as the density of electrons per unit volume, $D_{L(T)}$~ as the longitudinal (transverse) diffusion coefficient(s), $v_d$~ as the drift velocity and $\eta$~ as the attachment coefficient one can write:

\begin{eqnarray}
&\mathcal{N}_e(x',y',z',t)\! & = \! \frac{e^{-\frac{(x'-x)^2+(y'-y)^2}{4D_T(t-t_0)}} e^{-\frac{((z'-z)+v_d(t-t_0))^2}{4D_L(t-t_0)}}}{(4\pi D_T (t-t_0))(4\pi D_L (t-t_0))^{1/2}} \times \nonumber\\
&& \bar{n}_e \cdot e^{-\eta v_d (t-t_0)}\label{THESOL}
\end{eqnarray}
where the formula applies far from the TPC boundaries \citep{Gonzalez-Diaz:2017gxo}. Here $x'$, $y'$, $z'$ and $t$ denote position and time measured typically at the charge collection plane (the anode), and $(x,y,z,t_0)$ refer to the initial position and time of the ionization cloud, \eg\ the interaction point, assumed to be point-like and containing
$\bar{n}_e$ electrons. The solution in eq. \ref{THESOL} is an asymmetric gaussian cloud that broadens and losses carriers as it goes on. Arbitrary track topologies can be propagated directly by superposition of such solutions.
Choosing $z = 0$~as the coordinate of the amplification plane, and assuming that all charge arrives at a fixed time
$t$-$t_0 \simeq z/v_d$, eq.  \ref{THESOL} simplifies to:

\begin{eqnarray}
&\mathcal{N}_e(x',y',z') &=  \frac{e^{-\frac{1}{2}(\frac{x'-x}{D_T^*\sqrt{z}})^2} e^{-\frac{1}{2}(\frac{y'-y}{D_T^*\sqrt{z}})^2} e^{-\frac{1}{2}(\frac{z'}{D_L^*\sqrt{z}})^2}}{(2\pi D_T^{*,2} z)(2\pi D_L^{*,2} z)^{1/2}}\times \nonumber\\
&& \bar{n}_e \cdot e^{-\eta z}\label{THESOL3}
\end{eqnarray}
where $D_{L,T}^* = \sqrt{2D_{L,T}/v_d}$. As it turns out, the electrons produced in a \bbonu\ decay are extended track in a HPXe detector. One can still make use of  eq. \ref{THESOL3} making the approximation that the track consists in a superposition of point-like energy depositions that arrive at successive times to the anode.

Diffusion in pure xenon is large, with $D_L^* \sim {\rm 10~mm}/\sqrt{L}$, and
$D_T^* \sim {\rm 3~mm}/\sqrt{L}$, where $L$~is the drift length in meters. As a consequence, the ionization electrons produced by tracks located at relatively long distances from the anode will have spread considerably both in the longitudinal and transverse coordinates, and the resulting reconstructed image of the track will be consequently blurred. This undesirable effect can be limited by using mixtures that add a quencher capable of cooling the diffusion electrons and therefore reduce the diffusion.
%Alas, all known quenchers also absorb the scintillation light emitted by xenon, introducing the problems discussed above. The problem and potential solutions is discussed in section XX.

\subsection{Electron lifetime}

As primary electrons drift to the anode some of them will be absorbed by impurities in the gas. This results in the so-called ``electron lifetime'':
\begin{equation}
\tau_e = (\eta ~ v_d)^{-1}
\end{equation}

Thus, if the initial number of drift electrons is $n_0$ (at $t_0$), the number $n$~ reaching the anode at time $t$~will be:
\begin{equation}
n = n_0 \, e^{\frac{-(t-t_0)}{\tau_e}} = n_0 \,  e^{-\eta z}
\end{equation}

The main cause of attachment in large TPCs is related to the presence of O$_2$. For a given fraction of oxygen concentration, $f_{\textrm{O}_2}$, the lifetime is inversely proportional to both the square of the pressure, $P$ and $f_{\textrm{O}_2}, \tau_e \propto \frac{1}{P^2} \frac{1}{f_{\textrm{O}_2}}$.
%
%\begin{equation}
%\tau_e \propto \frac{1}{P^2} \frac{1}{f_{O_2}}
%\end{equation}
%
The only realistic way to achieve the necessary purity levels is through material selection and continuous recirculation and purification, to minimize the factor $f_{\textrm{O}_2}$. The dependence with $1/P^2$~ is also a major constraint for the operational pressure. In pure xenon the drift velocity is \MCDriftVelocity, and thus \SI{1}{\ms} is required to drift \SI{1}{\meter}. The drift length of a next-generation HPXe detector with a mass in the ton scale will be in the range \SIrange{1}{3}{\meter}. An electron lifetime in the range
 \SIrange{10}{30}{\ms} would translate in a
$10\%$ charge loss at the maximum drift length. This energy loss can be computed event by event, if \tz\ is known, and therefore a correction can be applied, with a residual relative error in the energy equal to the relative error in the determination of the lifetime. Thus, a relative error of $5\%$ in the determination of the lifetime would translate into a residual of $0.5\%$, which is of the same order of the practical intrinsic resolution in an electroluminescent HPXe TPC (see section \ref{sec.next}). Thus, long lifetimes  and precise lifetime corrections (which imply a measurement of \tz) are a must for a HPXe TPC aiming for the best energy resolution. Without \tz\ the fluctuations in energy introduced by attachment at the large drift distances become very large.

\subsection{Intrinsic energy resolution in xenon}
\begin{figure}[h!]
\centering
\includegraphics[width=1.0\textwidth]{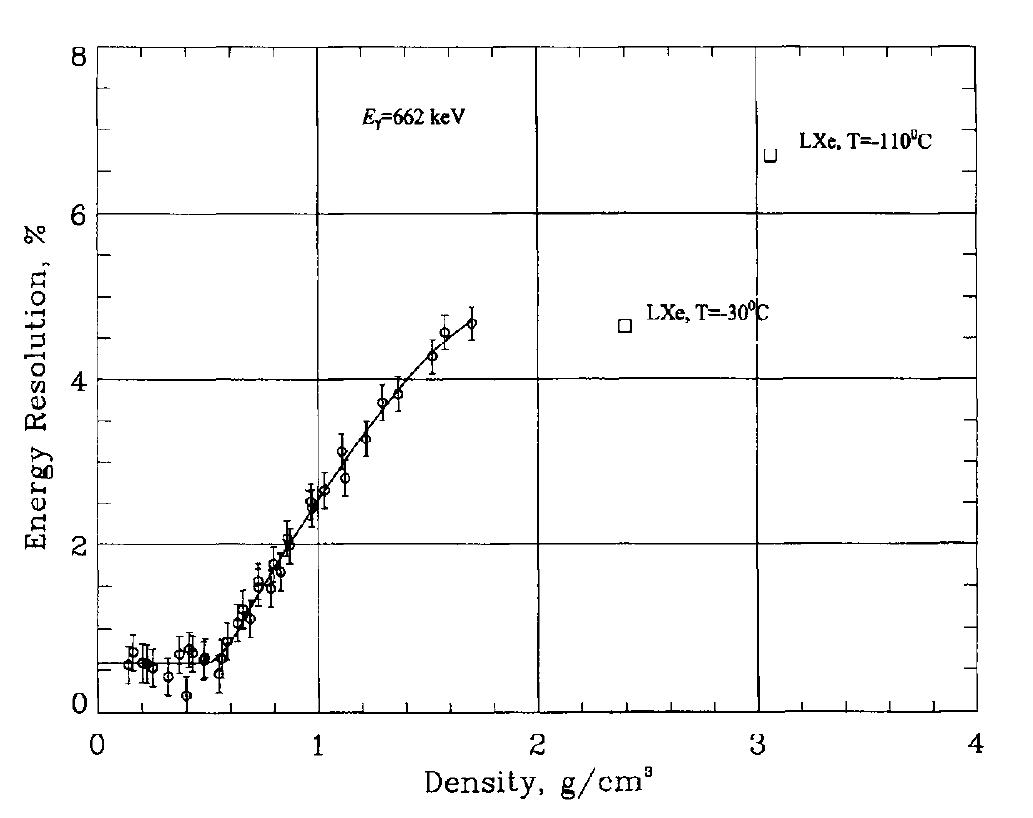} 
\caption{The energy resolution (FWHM) is shown for \ensuremath{{}^{137}\rm Cs} \SI{662}{keV} gamma rays, as a function of xenon density, for the ionization signal only. Reproduced from \citep{Bolotnikov:97}.} \label{fig:bolotnikov} 
\end{figure}

%%%
Excellent energy resolution is a crucial ingredient for a  \bbonu\ experiment. Indeed, physics allows such resolution to be attained in a gaseous xenon TPC. Instead, those very same physics processes limit the resolution in a liquid xenon TPC. This is clearly seen in \fig\ \ref{fig:bolotnikov}, reproduced from \citep{Bolotnikov:97}. The resolutions displayed were extracted from the photo-conversion peak of the 662 keV gamma ray from the $^{137}$Cs isotope. Only the ionization signal was detected. A striking feature in figure~\ref{fig:bolotnikov} is the apparent transition at density
$\rho_t \sim \SI{0.55}{\mbox{g/cm}^3}$. Below this density, the energy resolution is approximately constant:
\begin{equation}
\delta E/E = 6 \times 10^{-3} {\rm ~FWHM}.
\label{eq:intrinsic}
\end{equation}
For densities greater than $\rho_t$, energy resolution deteriorates rapidly, approaching a plateau at LXe density.

%%%%
%\begin{figure}[tbh]
%\centering
%\includegraphics[width=0.5\textwidth]{img/energy-resolution-density.jpg}
%\caption{The energy resolution (FWHM) is shown for $^{137}$Cs 662 keV gamma rays, as a function of xenon density, for the ionization signal only. Reproduced from \citep{Bolotnikov:97}.} \label{fig:bolotnikov}
%\end{figure}
%%%

The most plausible explanation underlying this strange behavior is the appearance, as density increases, of two-phase xenon (\citep{Nygren:2007zz} and references therein). In contrast, given the xenon critical density, the intrinsic resolution in the gas phase is very good up to pressures in the vicinity of \SI{50}{\bar}.
Extrapolating the observed relative resolution in figure \ref{fig:bolotnikov} as $1/\sqrt{E}$ to the \XE\ Q-value (\Qbb), allows to predict the intrinsic energy resolution in xenon gas at the region of
interest for \bbonu\ searches, to be $\delta E/E = 3 \times 10^{-3} {\rm ~FWHM}$.
%
%\begin{equation}
%\delta E/E = 3 \times 10^{-3} {\rm ~FWHM}.
%\label{eq:res}
%\end{equation}

Based on ionization signals only, the above energy resolution reflects an order of magnitude improvement relative to liquid xenon. For densities less than $\rho_t$, the measured energy resolution in \fig\ \ref{fig:bolotnikov} matches the prediction based on Fano's theory \citep{PhysRev.72.26}. The Fano factor $F$ reflects a constraint, for a fixed energy deposited, on the fluctuations of energy partition between excitation and the ionization yield $N_I$. For electrons depositing a fixed energy $E$, the rms fluctuations $\sigma_I$~ in the total number of free electrons $N_I$ can be expressed as $\sigma_I = \sqrt{F\ N_I}.$
%\begin{equation}
%\sigma_I = \sqrt{F\ N_I}.
%\label{eq:fano}
%\end{equation}
%
For pure gaseous xenon (GXe)  various measurements \citep{Nygren:2007zz} show that
$F_{\rm GXe} = 0.15 \pm 0.02$.
%\begin{equation}
%	F_{GXe} = 0.15 \pm 0.02
%\label{eq:gas-fano}
%\end{equation}
%
In liquid xenon (LXe), however, the anomalously large fluctuations in the partitioning of energy to
ionization produce an anomalous Fano factor of $F_{\rm LXe} \sim 20$,
%\begin{equation}
%	F_{LXe} \sim 20,
%	\label{eq:liquid-fano}
%\end{equation}
larger than the one corresponding to xenon gas by about two orders of magnitude.

\subsection{Electroluminescence}

\subsubsection{The Gas Proportional Scintillation Chamber}
\label{sec.GPSC}

%\begin{figure}[tbhp!]
%\begin{center}
%\includegraphics[width=0.5\textwidth]{img/GSC.pdf}
%%\includegraphics[width=0.8\textwidth]{img/PC.pdf}
%\end{center}
%\caption{\small Principle of a Gas Proportional Scintillation Counter. Reproduced from \citep{Conde:04}).}
%\label{fig:GPSC}
%\end{figure}

The principle of a Gas Proportional Scintillation Chamber (GPSC) is the following \citep{Conde:67}. An X-ray enters through the chamber window and is absorbed in a region of weak electric field
($>0.8~{\rm kV ~cm^{-1}~ bar^{-1}}$) known as the drift region. The ionization electrons drift under such field to a region of moderately high electric field (around $3 ~{\rm kV ~cm^{-1}~ bar^{-1}}$), the so-called scintillation
or EL region. There, each electron is accelerated so that it excites, but
does not ionize, the gas atoms/molecules.
The excited atoms decay, emitting UV light (the so-called
secondary scintillation), which is detected by photosensors. The intensity of the secondary scintillation light is at least two orders of magnitude stronger than
that of the primary scintillation. However, since
the secondary scintillation is produced while the
electrons drift, its latency is much longer than that for the primary scintillation, and its rise time is much longer
(up to hundreds of $\mu$s for \bbonu\ events, compared to a few ns). For properly chosen electric field strengths and EL region spatial widths, the number
$n_{ph}$ of secondary scintillation photons produced by
a single primary electron is nearly constant and can
reach values of the order of thousand photons per electron.
The average total number, $N_t$, of secondary
scintillation photons produced by an interaction
is then $N_t = n_{ph}\cdot N_I$, (recall that $N_I$~is the number of primary
ionization electrons) so the photosensor signal
amplitude is nearly proportional to E, hence the
name of gas proportional scintillation counter
(GPSC) for this device.

What made GPSCs extraordinarily attractive was their improved energy resolution compared with conventional Proportional Chambers (PC). In a PC
the primary electrons are made to drift
towards a strong electric field region, usually
in the vicinity of a small diameter (typically
\SI{25}{\mu m}) anode wire. In this region,
electrons engage in ionizing collisions that lead to
an avalanche with an average multiplication gain
$M$ of the order of $10^3$~ to $10^4$. If $M$ is not too
large, space charge effects can be neglected, and
the average number of electrons at the end of the
avalanche, $N_a = M\cdot N_I$, is also proportional to
the energy $E$ of the absorbed radiation (hence
the name proportional (ionization) counter given
to this device).
 However, for PC detectors, there are
fluctuations not only in $N_I$ but also in $M$; for GPSCs, since
the gain is achieved through a scintillation process
with almost no fluctuations, only fluctuations in
$N_I$ and in the photosensor need to be considered. Thus, a better energy
resolution can be achieved in the latter case.

\subsubsection{Electroluminescent yield}

A detailed Monte Carlo study of the energy resolution that can be achieved in a high pressure xenon TPC with electroluminescent amplification (HPXe-EL TPC) as a function of the EL yield
was performed in \citep{Oliveira:2011xk}. The study obtained a formula for the
the reduced electroluminescence yield, $\left(\frac{Y}{p}\right)$, as a function of the reduced electric field, $\left(\frac{E}{p}\right)$.
\begin{equation}
\left(\frac{Y}{p}\right)=\left(130\pm1\right)\left(\frac{E}{p}\right)-\left(80\pm3\right) \left[\textrm{photons electron}^{-1}\textrm{ cm}^{-1}\textrm{ bar}^{-1}\right]
\label{eq:yopfit}
\end{equation}
where
the reduced electroluminescence yield is defined as the number of photons emitted per primary electron and per unit of drift length divided by the pressure of the gas, and $E/p$~is expressed in ${\rm kV cm^{-1} bar^{-1}}$.
% The behavior of $\left(\frac{Y}{p}\right)$ with $\left(\frac{E}{p}\right)$ is approximately linear even when the actual ionization threshold is achieved at $\left(\frac{E}{p}\right)\sim3~\textrm{kV~cm}^{-1}\textrm{bar}^{-1}$. In Figure \ref{fig:j} it can be easily seen that this threshold is achieved since, for higher values of the electric field, the fluctuations in the secondary charge production, which are bigger than in the electroluminescence, start to dominate. The EL yield keeps its linear behavior while the probability of ionization is low.
%
%Performing a linear fit to the obtained points gave:
%
%where $E/p$~is expressed in ${\rm kV cm^{-1} bar{-1}}$.
The formula was found to be in good agreement with experimental data
measured at 1 bar \citep{Monteiro:2007}.
\begin{equation}
\left(\frac{Y}{p}\right)=140\left(\frac{E}{p}\right)-116 \left[\textrm{photons electron}^{-1}\textrm{ cm}^{-1}\textrm{ bar}^{-1}\right]
\label{eq:yodata}
\end{equation}

\subsubsection{Energy resolution in an EL TPC}
\label{sec.elres}
One of the desirable features of a \HPXeEL\ TPC is its excellent intrinsic energy resolution,
due to the small value of the Fano factor in gaseous xenon and the small fluctuations of the EL yield.
Following \citep{Oliveira:2011xk}, the resolution $R_E$~(FWHM) of a \HPXeEL\ TPC can be written as:

\begin{equation}
 R_E = 2\sqrt{2 \log{2}} \,\, \sqrt{ \frac{\sigma_e^2}{\bar{N}_e^2}
 + \frac{1}{\bar{N}_e} (\frac{\sigma_{EL}^2}{\bar{N}_{EL}^2}) +
 \frac{\sigma_{ep}^2}{\bar{N}_{ep}^2} + \frac{1}{\bar{N}_{ep}} (\frac{\sigma_q}{\bar{G}_q})^2}
 \label{eq.res0}
\end{equation}
In equation \ref{eq.res0} the factor $2\sqrt{2 \ln{2}}$  corresponds to the relation between the FWHM and the standard deviation $\sigma$, of a given probability distribution
(${\rm FWHM = 2\sqrt{2 \ln{2}}\sigma \sim 2.35\sigma}$ ). The first term of the expression is related to fluctuations in the number of primary charges created per event, $N_e$,
the second to fluctuations in the number of EL photons produced per primary electron,
$N_{EL}$, the third reflects the variations in the number of photoelectrons extracted to the photosensor (\eg\ PMTs, SiPMs) per decay, $N_{ep}$, and the fourth the distribution in the photosensor single electron pulse height, $G_q$.

The primary charge fluctuations are described by the Fano factor, $F = \sigma_e^2/\bar{N}_e$.
The fluctuations associated with the electroluminescence production are described by the parameter $J$~ defined as the relative variance in the number of emitted VUV photons per primary electron, $J = \sigma_{EL}^2/\bar{N}_{EL}$. The conversion of VUV photons
into photoelectrons follows a Poisson distribution, and thus
$\sigma_{ep}^2 = \bar{N}_{ep}$. For PMTs, the fluctuations in the photoelectron
multiplication gain can be described by $(\frac{\sigma_q}{G_q})^2 = 1$ \citep{Oliveira:2011xk}. Taking into account the previous relation equation \ref{eq.res0} can be rewritten as:
\begin{equation}
 R_E = 2.35 \, \, \sqrt{ \frac{F}{\bar{N}_e}
 + \frac{1}{\bar{N}_e} (\frac{J}{\bar{N}_{EL}}) + \frac{2}{\bar{N}_{ep}}}
 \label{eq.res}
\end{equation}

The first term in equation \ref{eq.res}, $R(Fano)$, correspond to the intrinsic resolution in xenon, the second term, $R(EL)$~to the resolution associated to fluctuations in
electroluminescence and the third, $R(EP)$~to fluctuations in the number of photoelectrons produced in the photosensor plane per \bbonu\ decay, $\bar{N}_{ep}$, which can be obtained as:

\begin{equation}
 \bar{N}_{ep} = k\ \bar{N}_e\ \bar{N}_{EL},
\end{equation}
where $k$~is the fraction of EL photons produced per \bbonu\ decay that gives rise to the production of a photoelectron.

%\begin{figure}[bhtp!]
%\centering
%\includegraphics[width=0.6\textwidth]{pool/imgKr/resolution_15_bar.png}
%\caption{\small Energy resolution terms and EL yield characteristic of an \HPXeEL\ as a function of the reduced electric field for an EL gap of \NewTpcELGap\ a value of $k \sim \NewK$  and a pressure of \NewPressure.}
%\label{fig.yield}
%\end{figure}

\begin{figure}[h!]
\centering
\includegraphics[width=1.0\textwidth]{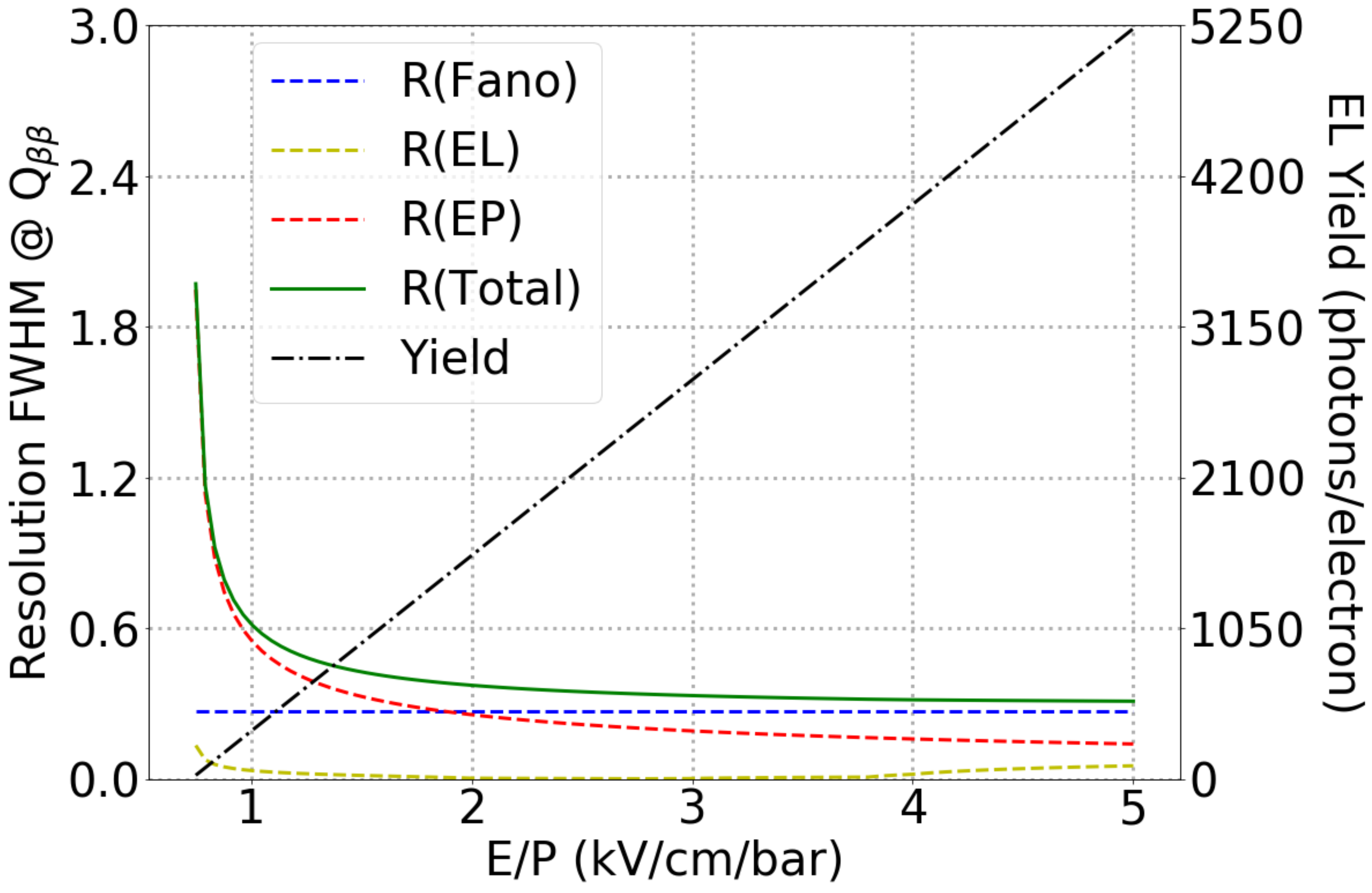}
\caption{\small Energy resolution terms and EL yield characteristic of a \HPXeEL\ TPC as a function of the reduced electric field for an EL gap of \NewTpcELGap\ a value of $k \sim \NewK$  and a pressure of \NewPressure.}
\label{fig.yield}
\end{figure}

Figure \ref{fig.yield} shows the EL yield and energy resolution, $R_E$ as a function of the reduced electric field of a \HPXeEL\ TPC operating at \NewPressure\ for the energy corresponding to \Qbb (\XenonQbb). The values of the EL width (relevant for the yield) and of $k$~(relevant for the resolution) correspond to those of the NEXT-100 detector. Notice that $R(Fano)$~is constant, $R(EL)$~is essentially negligible and $R(EP)$~improves quickly with the reduced field E/P. A typical value of operation for
E/P is such that $R(Fano) = R(EL)$ ($E/P \sim 2$). The combined resolution at this value
is around 0.5\% FWHM and the yield around $500$ photons per electron. Increasing $E/P$ results in very little resolution improvement.

%%%
%\begin{figure}[tb!]
%\centering
%\includegraphics[width=0.85\textwidth]{img/NextResolution}
%\caption{\small Pulse amplitude (open symbols) and energy resolution (full symbols) for 5.9 keV X-rays absorbed in an early NEXT prototype as a function of: E/p-scint, the reduced electric field in the scintillation region. Notice that the resolution for E/p above 3 is about the same for all pressures, near 8\%. This extrapolates to better than 0.5\% at \Qbb.}
%\label{fig.nr}
%\end{figure}
%
%Figure \ref{fig.nr} \citep{Fernandes:2010gg} presents experimental evidence of the intrinsic good resolution of a \HPXeEL, obtained with low energy x-rays.
%The figure shows pulse amplitude (open symbols) and energy resolution (full symbols) for 5.9 keV X-rays absorbed in a NEXT-0 prototype as a function of the reduced electric field in the scintillation region. Notice that the resolution for E/p above 3 is about the same for all pressures, near 8\%. This extrapolates to about 0.4\% at \Qbb.
%
%%%

\subsection{Avalanche multiplication in HPXe applied to \bbonu\ searches}

The use of avalanche multiplication in HPXe detectors searching for \bbonu\ processes finds two distinct problems. First,
the fluctuation in the gain, $G$, is considerably larger than $F$~in gas proportional counters involving avalanche multiplication, and thus becomes the dominant term in the resolution. Second, electron multiplication in pure xenon is difficult, due to the fact that VUV scintillation light, copiously produced with the multiplication process, ejects electron from metallic surfaces defining the electrodes. Those electrons, in turn, ionize the gas to the point of breakdown.

The use of quenchers, on the other hand, suppresses the primary scintillation light and has a heavy cost in terms of energy resolution and particle identification (without primary scintillation is not possible to define \tz). Two ways have been considered to solve this problem: a) the use of a ``magic gas'', capable to absorb the VUV light emitted by xenon and re-emit it at a more manageable wavelength (\eg\ in the visible region), without introducing extra fluctuations, and b) the use of micro-pattern devices, such as Micromegas  or GEMs whose confined geometrical structure makes them capable, a priory to operate at high pressure without quenchers.
%The possibility of operating an EL-HPXe detector with {\em homeopathic} amounts of quenchers will be discussed in the next section, where the performance and limitations of operating a gain HPXe with quenchers will also be presented.  On the other hand, the possibility of operating a pure xenon gas TPC with micro-pattern device, such as Micromegas has also been explored.

%\begin{figure}[tb!]
%\centering
%\includegraphics[width=0.6\textwidth]{img/MicromegasResolution.pdf}
%\caption{\small Resolution of a micro-bulk Micromegas as a function
%of the pressure for 22.1 keV photons. The resolution varies between 12\% at 1 bar ($\sim$ ) about 1\% at \Qbb, to
% 32\% at 10 bar, 3\% at \Qbb. The results were measured as a part of the NEXT collaboration R\&D
% \cite{Balan:2010kx}. }
%\label{fig.mm}
%\end{figure}

This second possibility was explored in \citep{Balan:2010kx}.
%and the results of the study
%are summarized in figure \ref{fig.mm}.
While Micromegas were found, indeed,
robust enough to allow operation in pure xenon and at high pressures, their resolution was measured to degrade with increased pressure. It was found that the resolution attainable at \Qbb\ by  micro-bulk micromegas at \SI{10}{\bar} would be $3\%$, to be compared with that of $0.4\%$ found in \citep{Fernandes:2010gg} using electroluminescence. Both measurements were carried out with very small setups, in close-to-ideal conditions, and therefore can be taken as reflecting the intrinsic performance of the devices under study.

No magic gas has been found so far capable to re-emit xenon scintillation light at longer wavelengths. Penning mixtures have been tried as a part of the R\&D of the NEXT collaboration, as will be further described in section \ref{sec.next}.

\subsection{Topological signature}
Another major advantage of gas relative to liquid ---and in general relative to high density calorimeters--- is the ability to exploit the topological signal of a \bbonu\ event, that is the capability to image the tracks left in the gas by the two electrons produced in the \bbonu\ decay. At
\SI{15}{\bar}
the track length of the electrons is of the order of \NextTrackLengthAtQbb\ and can be easily reconstructed in a HPXe TPC. Such a topological signature is not available in LXe detectors, due to the high density of the liquid phase ---in fact most other experimental techniques are based in high density calorimeters, none of which can reconstruct the electrons trajectory---.

\begin{figure}[h!]
\centering
\includegraphics[width=1.0\textwidth]{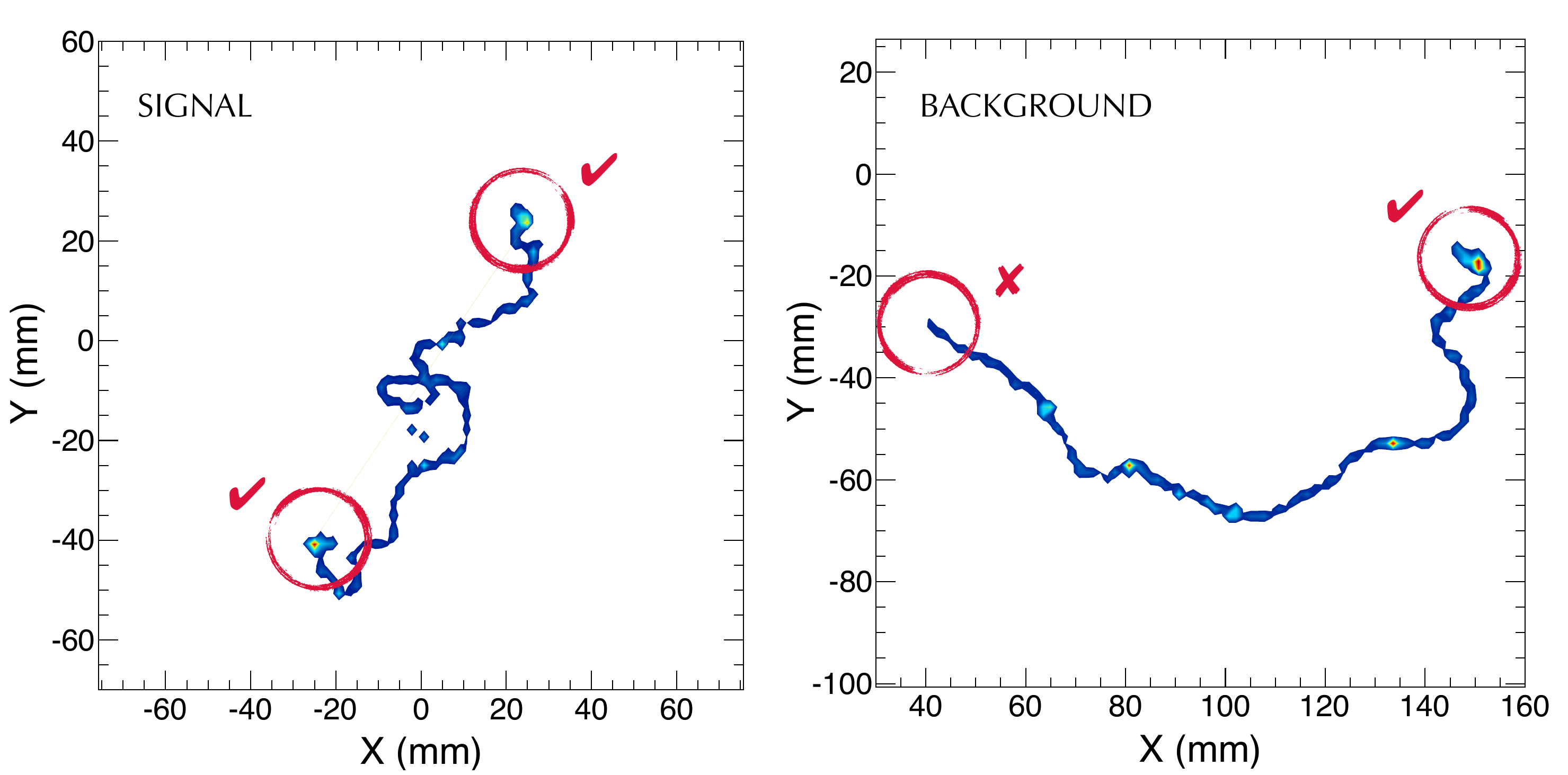}
\caption{\small The left panel shows two electrons emitted in a \bbonu\ decay propagating in HPXe with perfect track reconstruction; the right panel shows a single background electron produced by a photoelectric interaction from a \BI\ gamma of energy very close to \Qbb. While the energy of the background electron could enter the ROI, the topology of the latter is different from the former. A \bbonu\ event results in two electrons which are ended in two  \emph{blobs} of energy as the electron deposit suddenly its energy near the end-of-the-ionization path (Bragg peak). In the case of a background electron there is only a single blob. Figure from \citep{Martin-Albo:2015rhw}}
\label{fig:track} 
\end{figure}

An electron propagating in high density xenon ionizes the medium and results in a random trajectory due to large multiple scattering. Delta rays and bremsstrahlung photons are emitted along the trajectory. This is shown in \fig\ \ref{fig:track}. Even with a detector capable of reconstructing perfectly the electron trajectory, the track would still be wiggled and diffuse due to multiple scattering and the emission of delta rays and photons. Diffusion will further blur the ``electron picture''.

%%%
%\begin{figure}[htbp!]
%\centering
%\includegraphics[width=0.6\textwidth]{img2/topo.png}
%\caption{\small The left panel shows two electrons emitted in a \bbonu\ decay propagating in an HPXe with perfect track reconstruction; the right panel shows a single background electron produced by a photoelectric interaction from a \BI\ gamma of energy very close to \Qbb. While the energy of the background electron could enter the ROI, the topology of the later if different from the former. A \bbonu\ event results in two electrons which are ended in two  \emph{blobs} of energy as the electron deposit suddenly their energy near the end-of-the-ionization path (Bragg peak). In the case of a background electron there is only a single blob.}
%\label{fig:track}
%\end{figure}
%%%

%\begin{table}[tb]
%\begin{center}
%\begin{tabular}{cccc}
%\hline \hline
%E$_e$ (keV) & Probability (\%) & E$_{\gamma}$ (keV) & Mean path (cm) \\
%\hline
%500 &  4 & 12 & 0.14 \\
%800 &  6 & 27 &  0.8 \\
%1240 & 8 & 58 & 1.8 \\
%1680 &10 & 95 & 5.5 \\
%2000 & 11 &133 & 11\\
%2480 &14 &198 & 33\\
%\hline \hline
%\end{tabular}
%\end{center}
%\caption{\small Radiation probability for electrons in xenon at 10 bar. Average energy of the emitted photons and their mean free path in the HPXe is also shown.}
%\label{tab:brems}
%\end{table}

There are some differences between a background electron with energy near \Qbb\ and a ``double electron'' event emitted by a \bbonu\ decay where the \Qbb\ energy is shared between the two electrons. Naively one could expect to be able to distinguish the emission vertex, but this is not possible due to multiple scattering. Other expected features are: a) less ``satellite photons'' (\eg\ photons deposited in the chamber and not associated to the electron track) for signal than for background, due to the fact that the background electron has in average twice the energy of the signal electrons and radiates considerably more. Also, the initial trajectory of the background electron is less twisty than that of the double electrons, since multiple scattering is inversely proportional to the momentum. However, the more powerful ``smoking gun'' separating signal for background is the energy deposited at the end of the electron trajectory. Near the end of the ionization path (Bragg peak) the electrons deposit suddenly their remaining energy. One can define the
end-of-the-track blobs by two spheres of some suitable radius (typically of about
\SI{1}{cm}) around the end-points of the tracks. The energy contained in those spheres is the ``blob energy''. The energy of the lowest energy blob of single electrons originated in the gas by photoelectric or Compton interactions of gammas with energies close to \Qbb\ is much smaller than the energy of the highest energy blob.
%Single electrons originated in the gas by photoelectric or Compton interactions of gammas with energies close to \Qbb\ only have one blob, or to be precise the energy of the lowest energy blob is much smaller than the energy of the highest energy blob.
Instead, the energy of the two blobs is similar for electrons produced in double-electrons produced in a \bbonu\ decay.

%
%\begin{figure}[htbp!]
%\centering
%\includegraphics[width=0.6\textwidth]{img2/blob1_blob2.pdf}
%\caption{\small The left panel shows the energy of the blob of less energy versus the energy of the blob of high energy for a background electron propagating in an HPXe ---perfect track reconstruction---, while the right plot shows the same plot of the 2 electrons produced in a \bbonu\ propagating under the same conditions. In the first case the energy of the lower energy blob is much smaller than the energy of the higher energy blob, in the second case both are roughly the same.}
%\label{fig:bib2}
%\end{figure}

\begin{figure}[h!]
\centering
\includegraphics[width=0.49\textwidth]{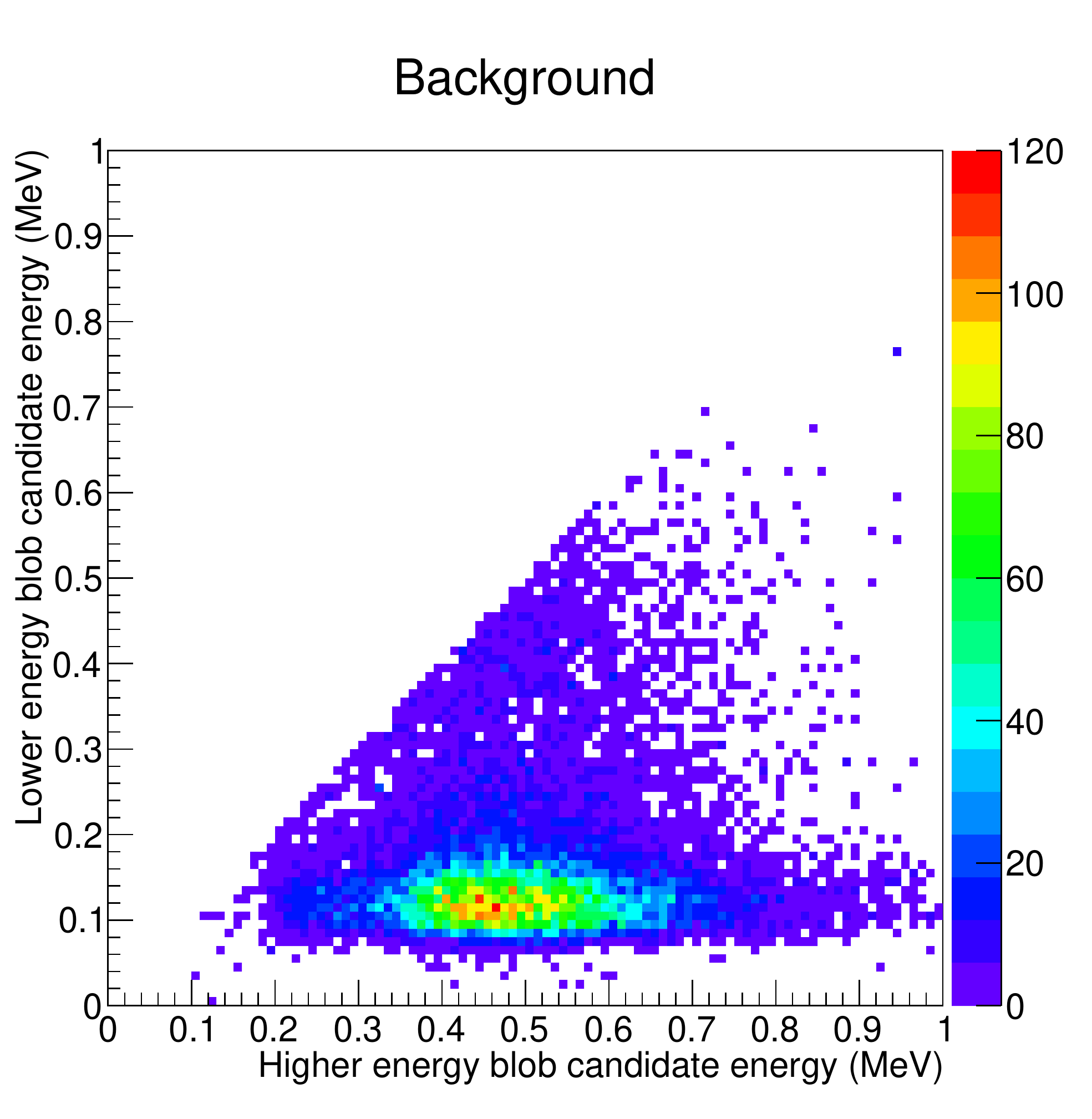}
\includegraphics[width=0.5\textwidth]{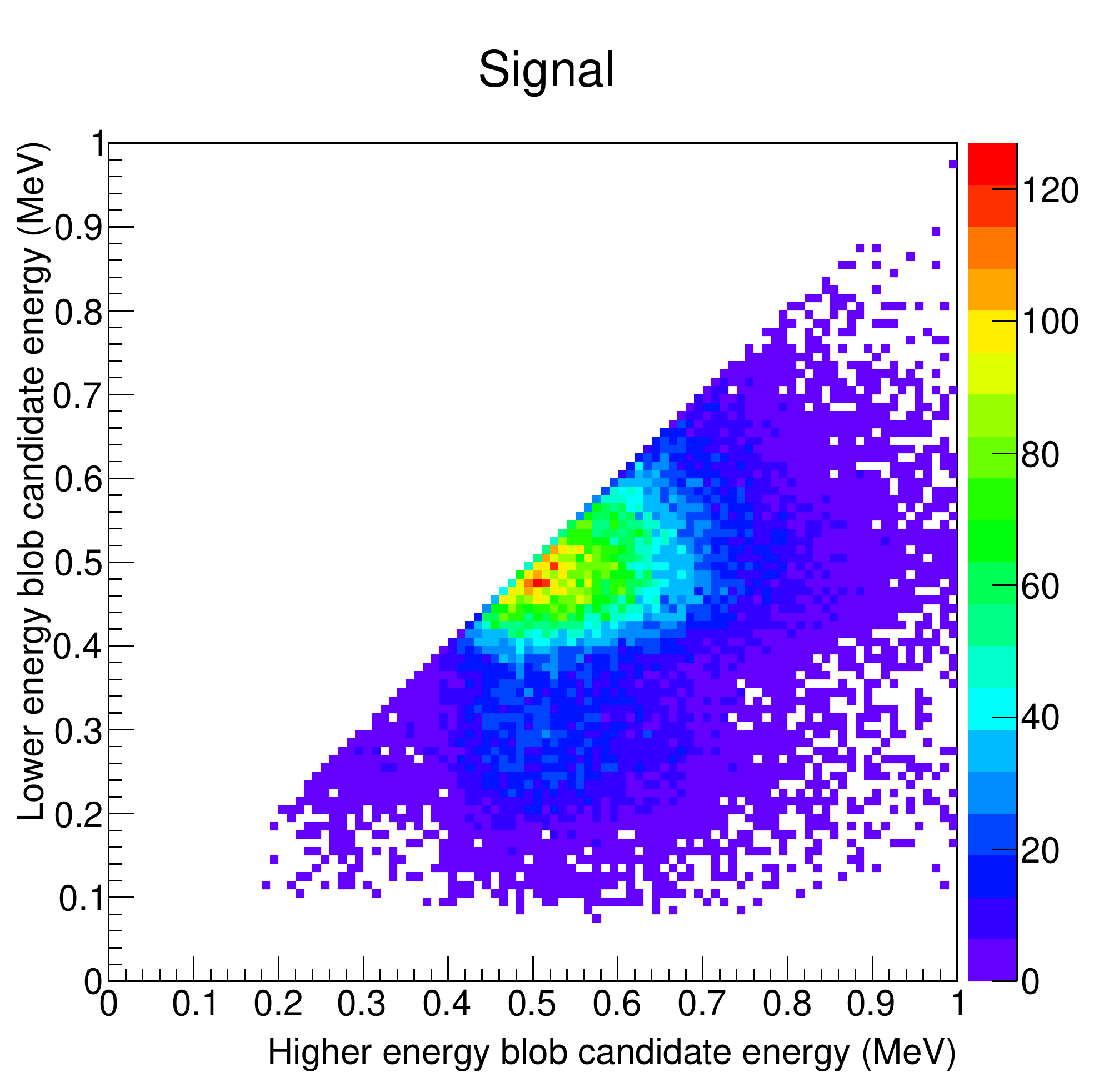}
\caption{\small The left panel shows the energy of the blob of less energy versus the energy of the blob of high energy for a background electron propagating in a HPXe TPC ---perfect track reconstruction---, while the right plot shows the same plot of the 2 electrons produced in a \bbonu\ propagating under the same conditions. In the first case the energy of the lower energy blob is much smaller than the energy of the higher energy blob, in the second case both are roughly the same.}
\label{fig:bib2} 
\end{figure}

Plotting the energy of each blob reveals a clear separation between signal and background event. The left panel of figure \ref{fig:bib2} shows the case for background electrons, while the right plot shows the case for signal. In the first case the energy of the lower energy blob is much smaller than the energy of the higher energy blob, in the second case both are roughly the same. A cut requiring that the energy of both blobs is larger than about
\SI{250}{keV} separates very effectively single and double electrons.

%However, the rejection capability of the topological signature goes beyond the separation between single and double electrons, as illustrated in \fig\ \ref{fig:geom_rejection}, where it can be seen that requesting a single track with no additional energy deposits (since those are likely associated to bremsstrahlung emission which happens more often for background electrons of energy \Qbb/2 than for signal events of average energy \Qbb/2) eliminates all potential backgrounds except those high energy electrons that manage to fake the double blob signature of the signal. We will discuss background rejection in an HPXeEL with more detail in section XX.

\section{The development of the HPXe and HPXe-EL technology for $\boldsymbol{\beta\beta 0\nu}$ searches}
\label{sec.history}

The first proposal to search for \bbonu\ decays using \XE\ was published in 1976 \citep{munari:1976} (an even earlier paper, by the same authors dates back to 1961
\citep{munari:1961}).  The proposed technology was a self-triggered cloud chamber filled with a mixture of xenon and helium, with helium being the permanent gas of the chamber and xenon acting as condensable vapor. Electroluminescence was proposed to trigger the chamber and to measure the energy of the particles. The setup included an electric field (to produce electroluminescence and clear the ionization) and a magnetic field whose role was to turn the tracks, and thus separate single electrons arising from backgrounds from double electrons arising from \bb\ decays. Thus, remarkably, two of the major assets of the HPXe-EL technology (\eg\, energy resolution thanks to proportional amplification of the ionization signal and a topological signature to distinguish two electrons from backgrounds) were identified in this pioneer work. 

The concept of gas proportional scintillation counter (GPSC), discussed in section \ref{sec.GPSC} dates from 1967 \citep{Conde:67}.  In 1975 the notion of GPSC was combined with that of a TPC, resulting in the Scintillation Drift Chamber (SDC) \citep{Charpak:75}.  A proposal to build a HPXe-EL TPC was made in 1983 \citep{Barabash:83} (see also \citep{Barabash:91}). A large SDC with 19 PMTs ~\citep{Bolozdynya:96} demonstrated excellent energy resolution at high pressure (9 bar) for high energy X-rays, consistently extrapolating
to $0.5\%$ FWHM at \Qbb. 

And yet, the early attempts to search for \bbonu\ processes with high pressure xenon chambers were not based in electroluminescence. 
The first such attempt was a small ionization chamber (with a mass of \SI{627}{\gram}) operating at a pressure of \SI{3}{\mega\pascal} (\SI{30}{\bar}) built at the Baksan neutrino observatory \citep{barabash:86}. The resolution achieved by this early apparatus ($2.7\%$ FWHM at \Qbb) was, in fact, better than that attained by subsequent detectors based in multiplication gain. The setup, however, had large \RAD\ contamination, due to radon emanating from the purification getters. The Baksan experiment managed to set a first limit of  \SI{5.5E19}{\yr} in the lifetime of the \bbonu\ decay for \XE.

At about the same time, a multi-element proportional chamber, operating at \SI{10}{\bar} and with a mass of around \SI{4}{\kg} was built by Fiorini and collaborators ---the so-called ``Milano experiment''--- \citep{Alessandrello:86}. The resolution achieved was $4.2\%$ FWHM at \Qbb. The setup was operated at the LNGS laboratory with natural xenon and with xenon enriched at 64\% in the isotope \XE. The detector was able to perform a crude reconstruction of the event topology that allowed the separation between ``single cluster events'' (\eg\ single or double electrons with no satellite energy depositions) and  ``multiple cluster events'' (\eg\ background events where additional energy deposition was identified). However the reconstruction could not separate between single and double electrons. The main source of contamination was the steel wires making up the cells as well as the titanium vessel. The background counting rates were 
$(5.1 \pm 0.3) \times 10^{-3}$~counts/keV/hour for the enriched xenon sample.
The best sensitivity achieved was \SI{1.2E22}{\yr} (at $95\%$ CL) in the lifetime of the \bbonu\ decay for \XE \citep{Alessandrello:88, Bellotti:89}. The experiment did not observe
the \bbtnu\ mode, setting up a limit of \SI{1.6E20}{\yr} at the $95\%$CL \citep{Bellotti:89}.

%Interestingly, the Milano experiment displayed many of the strengths of the HPXe technology, including the availability of a topological signature. The weak points, however, were a mediocre resolution and a high background rate, related with insufficient control of radioactive sources (\eg\ no internal shielding, radon, etc.). The counting rate was about
%45 events per keV and year. Taking a ROI of 1 FWHM ($\sim$ 100 keV), this corresponds to a total rate of 450 events per year for a detector of 5 kg. 

The St. Gotthard TPC (SGTPC) was built in 1987 \citep{Iqbal:1987vh}. It was a \SI{300}{\liter} HPXe detector, built with radiopure materials and operating at  a pressure of \SI{5}{\bar} in the St. Gotthard Tunnel Underground Laboratory. The experiment used xenon enriched at $62.5\%$ in the isotope \XE. Ionization was read by charge amplification in a wire plane and the transverse position was determined by XY strips. Being a TPC, the longitudinal position was determined by the arrival time of the ionization charge.

%Electron multiplication in pure xenon is difficult, due the fact that VUV scintillation light, copiously produced with the multiplication process ionizes in turn the gas to the point of breakdown. 

In order to quench the VUV light, the SGTPC used a mixture that contained 5\% of methane. As a consequence the TPC lacked the information about the start-of-the-event (\tz) and could not fiducialize the tracks along
the Z coordinate. The lack of \tz\ was a major limitation of this pioneer experiment, since one of the most important backgrounds for a HPXe TPC is the decay:
\begin{equation}
\BI \rightarrow \PO + e^- + \bar{\nu}_e {\rm ~(Q = 3.28 ~MeV, ~\tau_{1/2} = 19.7 ~min),}
\label{eq.bipo}
\end{equation}

Due to \RAD\ contamination, there is a steady-state concentration of \BI\ in the cathode. In a HPXe TPC with \tz\ the bismuth decay described in equation \ref{eq.bipo} can be vetoed by requiring a fiducial cut on Z. In the absence of \tz\ this background becomes dominant, although it can be partially vetoed by looking for a delayed coincidence with an alpha particle due to the decay:
\begin{equation}
\PO \rightarrow \PBD + \alpha  {\rm ~(Q = 7.8 ~MeV, ~\tau_{1/2} = 164 ~\mu s),}
\label{eq.bipo2}
\end{equation}

The energy resolution of the SGTPC was  6.6\% FWHM at \Qbb. This was way worse than the intrinsic resolution expected in xenon. The reasons for the degraded performance were the fluctuations in avalanche gain, and the quenching of the scintillation by the gas mixture, which introduced an irreducible additional fluctuation.

A first search for \bbonu\ events was published in 1991 \citep{Wong:1991vd}, followed by an improved result in 1998 \citep{Luscher:1998sd}. The total exposure of the experiment was \SI{12843}{\hour} or about \SI{1.5}{\yr}. The experiment set a limit of \SI{4.4E+23}{\yr} in the
lifetime of the \bbonu\ decay mode of \XE\ and a limit of \SI{0.72E+22}{\yr} in the
\bbtnu\ mode.

The SGTPC was the first detector to demonstrate the topological signature available to a (high pressure) gas TPC. A peculiarity of the experimental technique was that the scan of candidates was done visually. The efficiency and rejection power of the procedure was evaluated by mixing Monte Carlo events with real data. A $68\%$ acceptance for double electrons and a rejection power of $96.5\%$ for single electrons was found. This performance is similar to the best results obtained by the NEXT and PANDA-III-X collaborations using deep neural networks, as will be discussed in sections \ref{sec.next} and \ref{sec.axel}. 

The dominant source of background in the SGTPC were electrons emanating from the cathode that could not be rejected due to the lack of \tz. Still, the background rate at the ROI of the detector was $0.01$\ckky, the lowest among all the \bbonu\ searches of the time.  

%\begin{figure}[tbhp!]
%\begin{center}
%\includegraphics[width=0.8\textwidth]{img/EXOWP.pdf}
%\end{center}
%\caption{\small A conceptual drawing of the original proposal for the EXO detector, from the EXO white paper\footnote{available online at https://www-project.slac.stanford.edu/exo/docs/white_v5.pdf}. }
%\label{fig:EXO} 
%\end{figure}

The original proposal for the EXO detector, published in 1999\footnote{available online at \url{https://www-project.slac.stanford.edu/exo/docs/white_v5.pdf}} presented a large version of the
SGTPC. Like the former, it used multiplication gain to read the ionization, substituting the wire-pad arrangement by micro-pattern structures (GEMs). Operation at a pressure of \SI{5}{\bar} was assumed. PMTs placed in the barrel region were used to read the primary scintillation and thus measure \tz. In addition, a system of lasers to tag the \Bapp\ ion produced in the \bbonu\ decay of \XE\ {\em on the fly}, was envisioned.  

The EXO design assumed that the xenon would be mixed with another gas capable of quenching the xenon VUV light (so that the detector could be operated in the regime of charge multiplication without breaking the gas) and at the same time capable of neutralizing one of the positive charges of the \Bapp\ ion, so that tagging based in laser-induced resonance excitation of \Bap\ could be used. At the same time, this gas should be able to cool the drifting electrons, reducing diffusion and re-emit in the visible region, so that \tz\ could be measured.  
Alas, such a ``magic gas'' was not found neither by EXO, nor by an extensive R\&D conducted later by NEXT (see section \ref{sec.next}). 

%It is, indeed possible to find a mixture that reduces transversal diffusion without quenching the VUV light (\eg\ xenon-helium). Mixing with trace amounts of methane may reduce both longitudinal and transverse diffusion (at a rather large cost, however, for the primary scintillation). A mixture with a penning gas such as TMA, in turn, results in fast drift and minimal diffusion, and in addition transfers \Bapp\ to Ba$^+$, but completely quenches the VUV scintillation light with no usable re-emission. Thus, TMA mixtures, as studied by NEXT and later adopted by the PANDAX-III proposal, hit the same bottleneck found by the SGTPC, that is, lack of \tz. We will discuss quenchers in more detail in section XXX. 

To summarize, the GPSC was invented in 1965 and electroluminescence was proposed as early as 1975 for a \bbonu\ detector. However, the Baksan, Milano and St. Gotthard  experiments were all based in electron multiplication for the readout of the ionization. None of these experiments had good energy resolution, but the St. Gotthard TPC demonstrated a powerful topological signature and achieved a very good background rate in the ROI (for the time), in spite of the lack of \tz\ which was responsible for most of its background rate. 
The original EXO proposal also considered avalanche gain as the choice option and later switched to liquid xenon. 

The possibility of using a HPXe-EL TPC for \bbonu\ searches was resurrected in 2009
\citep{Nygren:2009zz}, and adopted as the baseline solution by the NEXT collaboration in its Letter of Intent \citep{Granena:2009it}. The choice of an asymmetric HPXe-EL TPC with an energy plane based in PMTs for energy measurement and a tracking plane based in SiPMs for tracking was established in the NEXT conceptual design report \citep{Alvarez:2011my} and further developed in the technical design report (TDR) \citep{Alvarez:2012haa}
%\footnote{The NEXT collaboration was formed in 2008 by Gomez-Cadenas. Nygren joined in 2009 and the late James White shortly after. The key ideas converging in the NEXT TDR were the use of electroluminescence, and the notion of separated functions for the energy and tracking plane, proposed by Nygren; the choice of an asymmetric detector, which permitted different sensor types for the energy and tracking plane, put forward by White; and the adoption of a tracking plane based in bare SiPMs acting as ``light pixels'', proposed by Gomez-Cadenas.}. 

\section{The NEXT program}
\label{sec.next}
The Neutrino Experiment with a Xenon TPC (NEXT) is an experimental program  developing the technology of high-pressure xenon gas Time Projection Chambers (TPCs) with electroluminescent amplification for neutrinoless double beta decay searches (\bbonu). 

The first phase of the program included the construction, commissioning and operation of two prototypes, called 
NEXT-DBDM and NEXT-DEMO (with masses around 1 kg). The \NEW\footnote{Named after the late Prof.~James White, a pioneer of the technique and a crucial scientist for the experiment.} detector, holding \SI{5}{kg} of xenon  implements the second phase of the program.  \NEW\ has been running successfully since October 2016 at Laboratorio Subterr\'aneo de Canfranc (LSC), Spain.  

\Next\ constitutes the third phase of the program. It is a radiopure detector deploying \SI{100}{kg} of xenon at \SI{15}{\bar} and scaling up \NEW\ by slightly more than 2:1 in longitudinal dimensions. In addition of a physics potential competitive with the best current experiments in the field, \Next\ can be considered as a large scale demonstrator of the suitability of the \HPXeEL\ technology for detector masses in the ton-scale. 

%A possible fourth phase of the program would be the \NHD\ detector. \NHD\ will improve the ``high-definition'' of the technology by reducing diffusion (using low-diffusion gas mixtures) and increasing the density of the tracking plane.

The fourth envisioned phase of the program is called  \Ntk, a detector that would multiply the mass of \Next\ by a factor \NtkTimesNext\ while at the same time reducing \Next\ background in the ROI by at least one order of magnitude, thanks to the combination of an improved topological signature and a reduced radioactive budget. Furthermore, \Ntk\ could implement \Bapp-tagging 
based in single molecule fluorescence imaging (SMFI) (see section \ref{sec.smfi}).
%whose first proof-of-concept has recently been published by the NEXT collaboration~\citep{McDonald:2017izm}. 
The importance of SMFI  \Bapp-tagging cannot be overemphasized, since it would permit {\bf a background free experiment at the ton scale} leading to a full exploration of the Inverse Hierarchy (IH) and beyond, with a high probability of a discovery. 

\subsection{The DEMO and DBDM prototypes}
\label{sbs.demo}

%\begin{figure}
%\centering
%\includegraphics[width=0.7\textwidth]{img/DemoSetup.jpg}
%\includegraphics[width=0.7\textwidth]{img/dbdm_vessel.jpg}
%\caption{\small The NEXT-DBDM detector. The NEXT-DEMO prototype setup at IFIC.} \label{fig.DEMO}
%\end{figure}
%
%

%\begin{figure}[h!]
%\centering
%\includegraphics[width=0.6\textwidth]{img/dbdm_resolution.pdf}
%\caption{\small Energy resolution measured by NEXT-DBDM: Data points show the measured energy resolution for \SI{662}{\keV} gammas (squares),
%$\sim$ \SI{30}{\keV} xenon X-rays (triangles) and LED light pulses (circles) as a function of the number of photons detected. The expected resolution including the intrinsic Fano factor, the statistical fluctuations in the number of detected photons and the PMT charge measurement variance is shown for X-rays (dot dot dashed) and for 662 keV gammas (dot dot dot dashed). Resolutions for the 662 keV peak were obtained from \SI{15}{\bar} data runs while X-ray resolutions we obtained from \SI{10}{\bar} runs. Figure from 
%\citep{Alvarez:2012hh}.}
%\label{fig.ERES} 
%\end{figure}

%
%%%%%%
%\begin{figure}
%\centering
%\includegraphics[width=0.8\textwidth]{img/TopoDemo.png}
%\includegraphics[width=0.8\textwidth]{img/dbdm_res.pdf}
%
%\caption{\small Left panel: the full energy spectrum measured for electrons of 511 keV in the DEMO detector. Right panel: the spectrum near the photoelectric peak for 662 keV electrons in NEXT-DBDM.}
%\label{fig.ERES}. 
%\end{figure}

The NEXT prototypes were the first HPXe-EL chambers built since the pioneer St. Gotthard TPC experiment discussed in section \ref{sec.history}. NEXT-DBDM was  instrumented with a single energy plane made of an array of 19 Hamamatsu R7378A 1'' photomultipliers capable to operate up to \SI{20}{\bar} pressure. The detector geometry was designed to minimize the dependence of light collection with position. Without a light 
sensor array near the EL region precise tracking information was not available and only coarse average position could be obtained using the PMT array light pattern. That was sufficient, nonetheless, to fiducialize events within regions of the TPC with uniform light collection efficiencies. 

\begin{figure}[h!]
\centering
\includegraphics[width=1.0\textwidth]{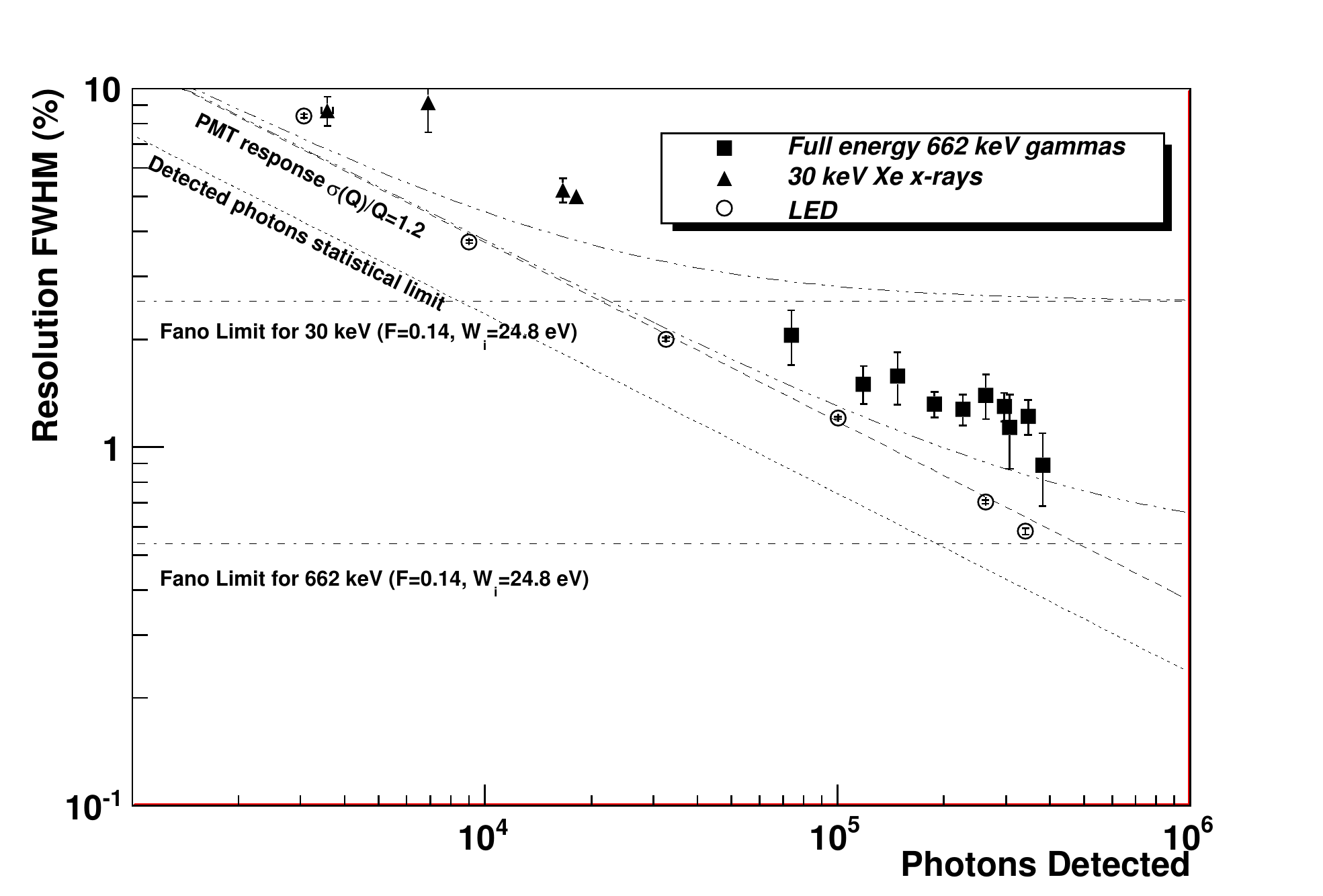}
\caption{\small Energy resolution measured by NEXT-DBDM: Data points show the measured energy resolution for \SI{662}{\keV} gammas (squares),
$\sim$ \SI{30}{\keV} xenon X-rays (triangles) and LED light pulses (circles) as a function of the number of photons detected. The expected resolution including the intrinsic Fano factor, the statistical fluctuations in the number of detected photons and the PMT charge measurement variance is shown for X-rays (dot dot dashed) and for \SI{662}{\keV} gammas (dot dot dot dashed). Resolutions for the  \SI{662}{\keV} peak were obtained from \SI{15}{\bar} data runs while X-ray resolutions we obtained from \SI{10}{\bar} runs. Figure from 
\citep{Alvarez:2012hh}.}
\label{fig.ERES} 
\end{figure}

Figure \ref{fig.ERES} shows the most important result obtained with DBDM. The energy resolution was measured with xenon X-rays and with a radioactive \CS\ source, at pressures of $10$ and \SI{15}{\bar}. The results obtained approached the intrinsic resolution that could be achieved in xenon (see discussion in section \ref{sec.elres}), extrapolating to 0.5\% FWHM at \Qbb\footnote{For all energy extrapolations in this report we use, unless otherwise stated, the simple statistical 1/$\sqrt{E}$ dependence.}. 

NEXT-DEMO was a larger-scale prototype of NEXT-100. The pressure vessel had a length of \SI{60}{\cm} and a diameter of \SI{30}{\cm}. The vessel could withstand a pressure of up to \SI{15}{\bar} but was normally operated at  \SI{10}{\bar}. The energy plane was instrumented with 19 Hamamatsu R7378A PMTs (same model than DBDM) and a tracking plane made of 256 Hamamatsu silicon photomultipliers (SiPMs). 
The detector operated for several years demonstrating: excellent operational stability, with no leaks and very few sparks; (b) good energy resolution ; (c) electron reconstruction using the SiPM tracking plane; (d) excellent electron drift lifetime, of the order of \SI{10}{ms} \citep{Alvarez:2012xda, Alvarez:2012hu, Alvarez:2013gxa, Alvarez:2012hh, Lorca:2014sra}. 

%\begin{figure}[h!]
%\centering
%\includegraphics[width=0.8\textwidth]{img/demo_topo.png}
%\caption{\small Energy distribution of the blobs at the end-point of single electrons coming from \NA\ decays (left) and tracks (mostly electron-positron pairs) coming from the \TL\ double escape peak (right). Figure from 
%\citep{Ferrario:2015kta}.}
%\label{fig.topo} 
%\end{figure}

\begin{figure}[h!]
\centering
\includegraphics[width=0.49\textwidth]{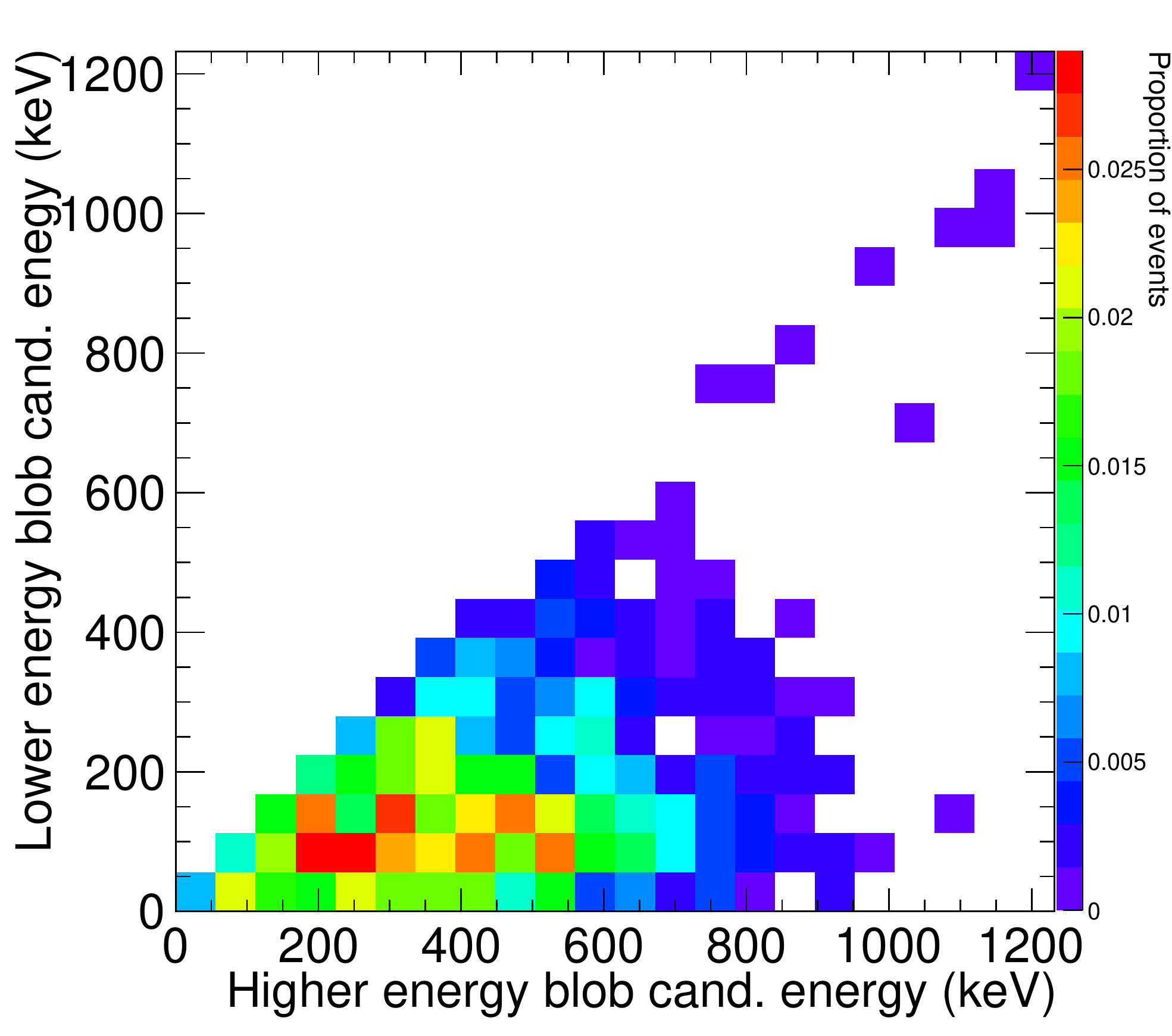}
\includegraphics[width=0.49\textwidth]{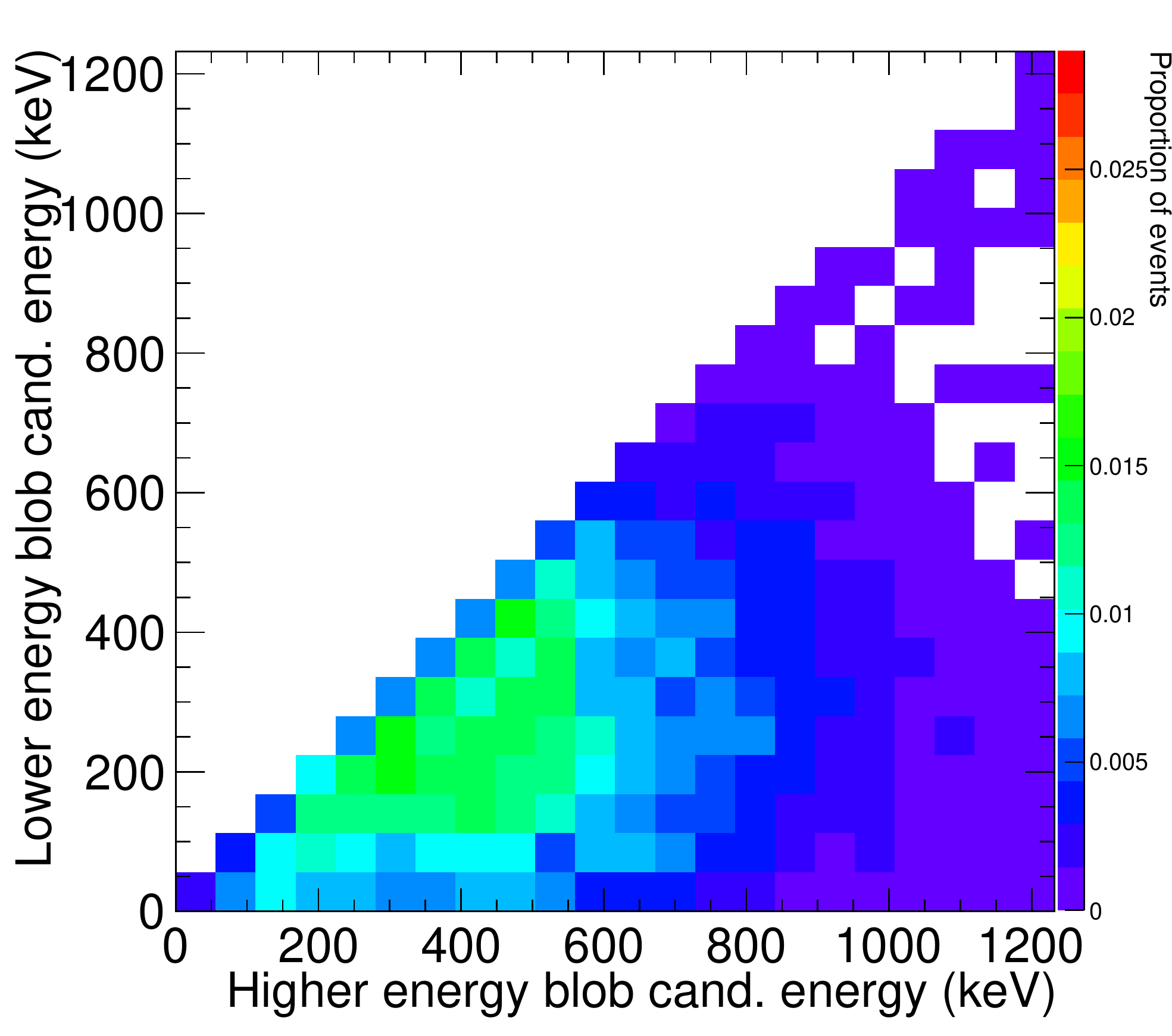}
\caption{\small Energy distribution of the blobs at the end-point of single electrons coming from \NA\ decays (left) and tracks (mostly electron-positron pairs) coming from the \TL\ double escape peak (right). Figure from 
\citep{Ferrario:2015kta}.}
\label{fig.topo} 
\end{figure}

Figure \ref{fig.topo} shows the first demonstration of topological signature in an
HPXe-EL TPC. The left panel corresponds to single electrons due to photoelectric interactions
of the \SI{511}{\keV} gamma emitted in \NA\ decays. The right panel corresponds to tracks selected in the double escape peak of \TL\ and thus it is enriched in ``double electrons'' (pairs electron-positron). The plots show clearly the difference between ``electrons-like'' and ``double-electrons like'' events, which can be easily separated with a cut on the energy of the lower-energy blob as extensively discussed in  \citep{Ferrario:2015kta}. 
%More discussion on the topological signature follows in the following sections. 

\subsection{The \NEW\ detector}
\label{sec.new}

%\begin{figure}[bhtp!]
%\centering
%\includegraphics[width=0.49\textwidth]{img/NEW.jpg}
%\includegraphics[width=0.49\textwidth]{img/newCollague.jpg}
%\caption{\small Left: the \NEW\ detector at the LSC. Right: a selection of the main subsystems of \NEW: a) the field cage; b) the anode plate; c) high voltage feedthrough; d) energy plane; e) PMTs used in the energy plane; f) tracking plane; g) kapton boards composing the tracking plane.} \label{fig.new2}
%\end{figure} 

%\begin{figure}[bhtp!]
%\centering
%\includegraphics[width=0.6\textwidth]{pool/imgNEW/NEW.png}
%\caption{\small The \NEW\ detector.} \label{fig.new}
%\end{figure} 
%
%\begin{figure}[bhtp!]
%\centering
%\includegraphics[width=0.6\textwidth]{pool/imgNEW/NEW.jpg}
%\caption{\small The \NEW\ detector at the LSC.} \label{fig.new2}
%\end{figure} 

\begin{figure}[h!]
\centering
\includegraphics[width=0.49\textwidth]{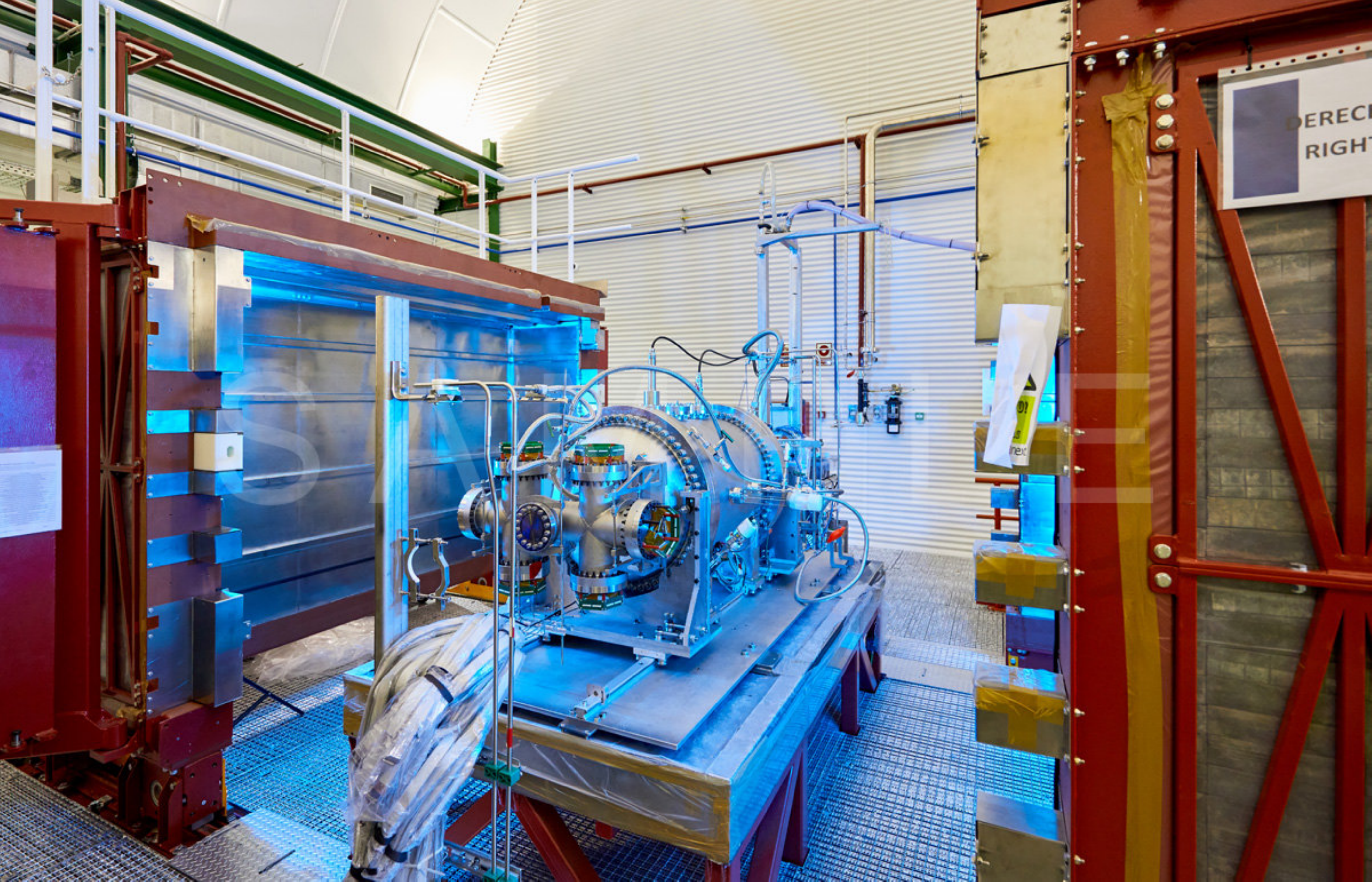}
\includegraphics[width=0.49\textwidth]{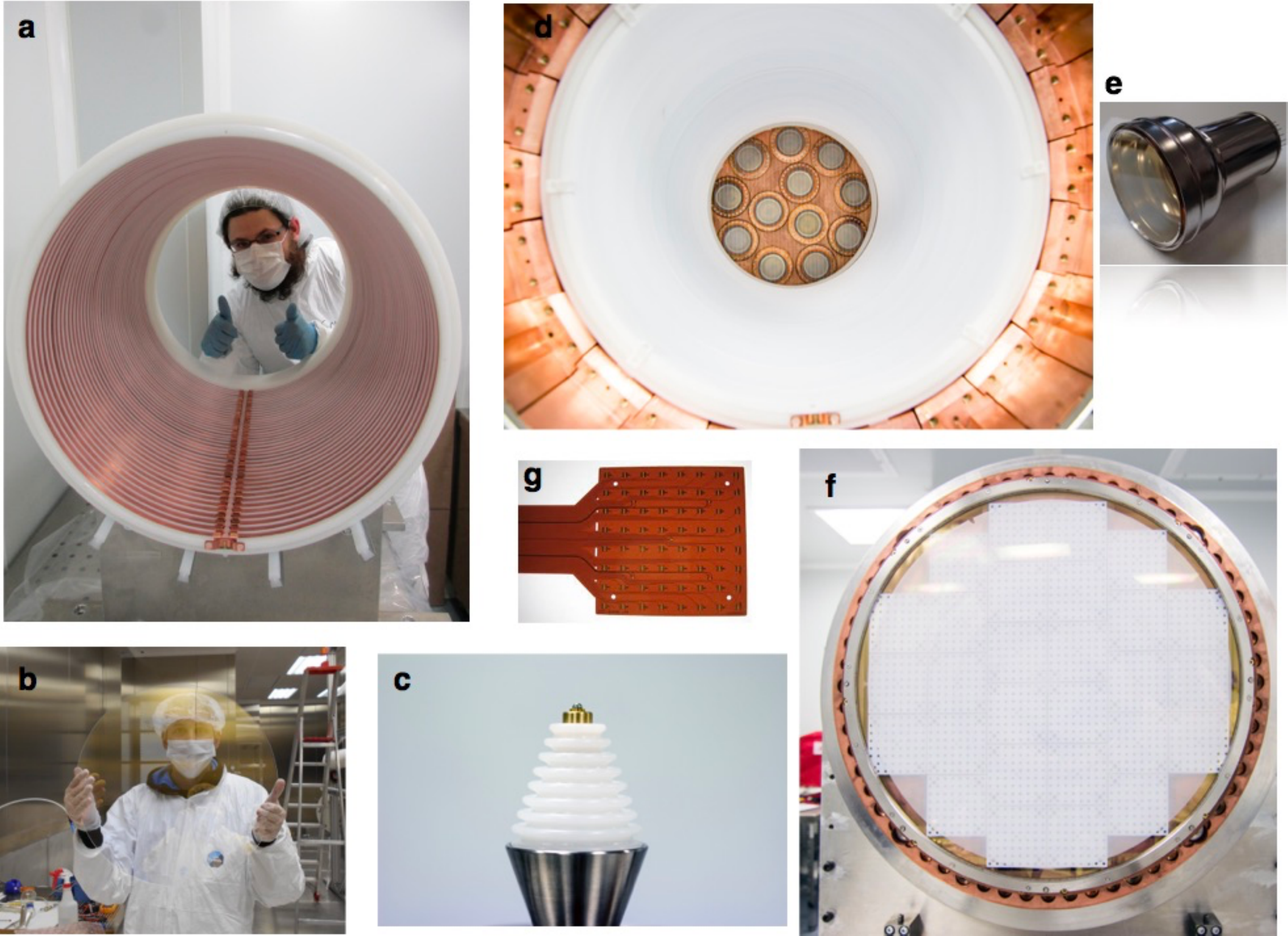}
\caption{\small Left: the \NEW\ detector at the LSC. Right: a selection of the main subsystems of \NEW: a) the field cage; b) the anode plate; c) high voltage feedthrough; d) energy plane; e) PMTs used in the energy plane; f) tracking plane; g) kapton boards composing the tracking plane.} \label{fig.new2}
\end{figure}

The \NEW\ apparatus ~\citep{Monrabal:2018xlr}, shown in figure \ref{fig.new2}, has roughly the same dimensions as the St. Gotthard TPC experiment, and is currently the world's largest \HPXeEL\ TPC.  
 The detector operates inside a pressure vessel fabricated with a radiopure titanium alloy, \NewPressureVesselMaterial. The pressure vessel sits on a seismic table and is surrounded by a lead shield (the lead castle). Since a long electron lifetime is a must, the xenon circulates and is purified in a gas system described in great detail in \citep{Monrabal:2018xlr}. The whole setup sits on top of a tramex platform elevated over the ground at Hall-A, in the LSC. 

 The right panel of figure \ref{fig.new2} shows a selection of the main subsystems of \NEW. The TPC  is, in essence, a scaled-down version (2:1) of the NEXT-100 TPC, and its construction and operation has been essential to guide the design of the latter. 
It has a length of \NewTpcLength\ and a diameter of \NewTpcDiameter. The field cage body is a High Density Polyethylene (HDPE) cylindrical shell of \NewFieldCageHDPEThickness\ thickness. The inner part of the field cage body is machined to produce grooves where radiopure copper rings are inserted (figure \ref{fig.new2}-a). The drift field is created by applying a voltage difference between the cathode and the gate, through high voltage feedthroughs (figure \ref{fig.new2}-c). The field transports ionization electrons to the anode where they are amplified. The drift length is  \NewTpcDriftLength, and the drift voltage is \NewDriftField.

The amplification or electroluminescent region is the most delicate part of the detector, given the requirements for a high and yet very uniform electric field. The anode is defined by a \PDOT\ (PEDOT) surface coated over a fused silica plate of \NewAnodePlateDiameter\ diameter and \NewAnodePlateThickness\ thickness (figure \ref{fig.new2}-b). The entire region is mounted on top of the tracking plane to ensure its flatness and is only connected to the rest of the field cage when closing the detector. A thin layer of \TPB\ (TPB), commonly used in noble gases detectors to shift VUV light to the visible spectrum 
%\cite{Gehman:2011xm}
is vacuum-deposited on top of the PEDOT. The EL gap is \NewTpcELGap\ wide.

The measurement of the event energy as well as the detection of the primary scintillation signal that determines the \tz\ of the event is performed by the \NEW\ energy plane (EP), shown in figure \ref{fig.new2}-d. The \NewTypePMT\ PMTs (figure \ref{fig.new2}-e) are chosen for their low radioactivity (\NewPMTActivity in \ensuremath{^{214}}Bi) \citep{Cebrian:2017jzb} and good performance.
%\cite{HamamatsuPMTs}). 
Since they cannot withstand high pressure they are protected from the main gas volume by a radiopure copper plate, \NewPmtEndCapThickness\ thick, which also acts as a shielding against external radiation. The PMTs are coupled to the xenon gas volume through \NewNumberOfPMT\ sapphire windows welded to a radiopure copper frame that seals against the copper plate.
The windows are coated with PEDOT %\cite{sigma:pedot}
in order to define a ground while at the same time avoiding sharp electric field components near the PMT windows. A thin layer of TPB is vacuum-deposited on top of the PEDOT.

The tracking function in \NEW\ is performed by a plane holding a sparse matrix of SiPMs. The sensors have a size of \NewSiPMSize\ and  are placed at a pitch of \NewSipmPitch. The tracking plane is placed \TrackingPlaneToAnode\ behind the end of the quartz plate that defines the anode with a total distance to the center of the EL region of \TrackingPlaneToEL. The sensors are \NewSiPMSeries\ series model \NewSiPMModel with \NewSipmCell\ cell size and a dark count of less than \NewSipmDarkCount\ at room temperature.  The cell's size is sufficient to guarantee good linearity and to avoid saturation in the expected operating regime ($\sim$ \NewPhotoelectronsPerSiPM). The SiPMs are distributed in \NewNumberOfBoards\ boards (DICE boards) with \NewNumberOfSiPMPerBoard\ pixels each for a total of \NewNumberOfSiPM\ sensors (figure \ref{fig.new2}-g). The DICE boards are mounted on a \NewTrackingPlaneEndCapThickness\ thick copper plate intended to shield against external radiation. The material used for the DICE boards is a low-radioactivity kapton printed circuit with a flexible pigtail that passes through the copper where it is connected to another kapton cable that brings the signal up to the feed-through. Each DICE has a temperature sensor to monitor the temperature of the gas and SiPMs and also a blue LED to allow calibration of the PMTs at the opposite end of the detector. The \NEW\ tracking plane (figure \ref{fig.new2}-f) is currently the only large system deploying SiPMs as light pixels in the world.
%
%\begin{figure}[!htb]
%\centering
%\includegraphics[angle=0, width=0.7\textwidth]{img/new_results.png}
%\caption{A selection of the results obtained by \NEW. See text for details.}
%\label{fig.newd}
%\end{figure}

\begin{figure}[h!]
\centering
\includegraphics[width=0.46\textwidth]{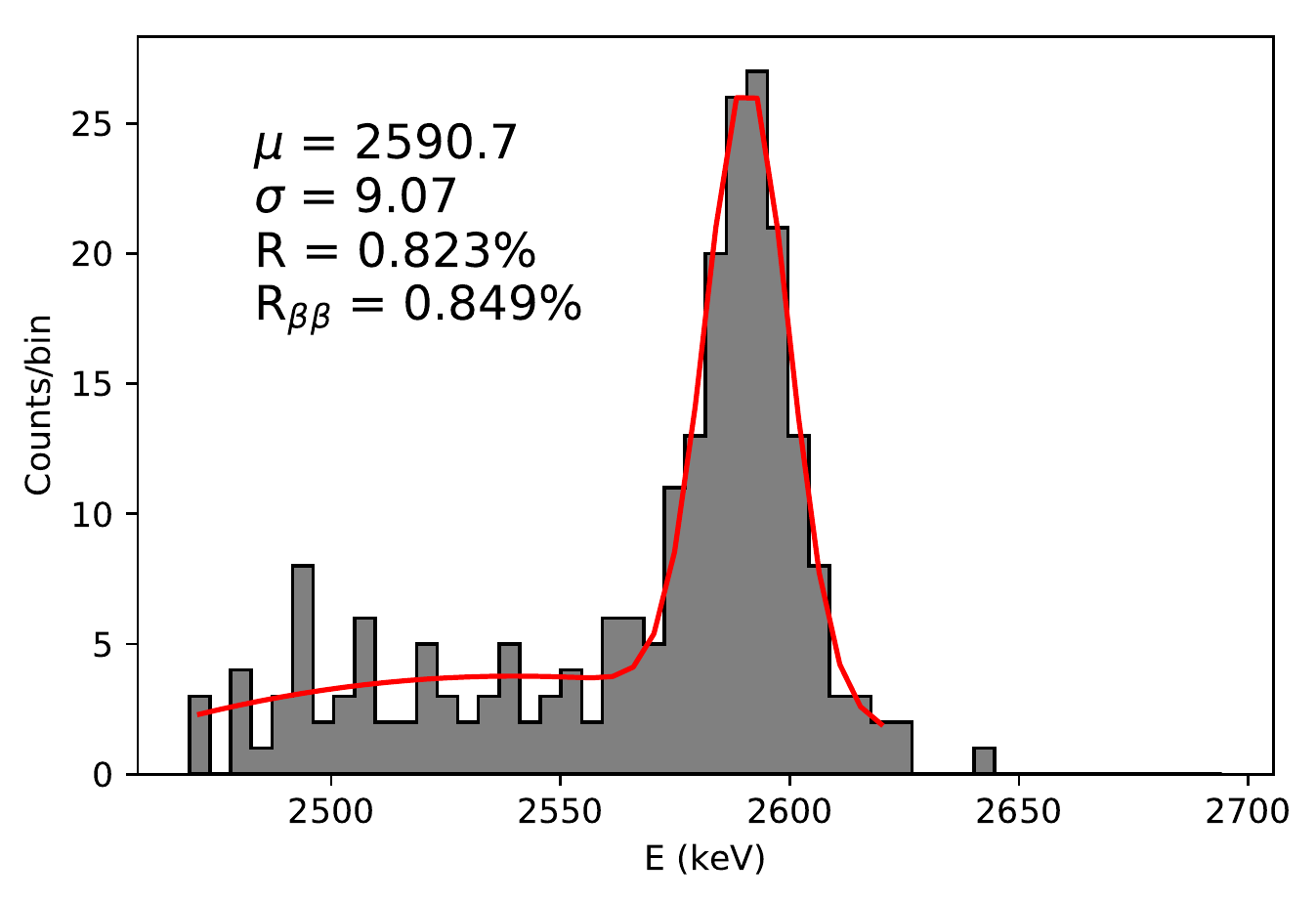}
\includegraphics[width=0.49\textwidth]{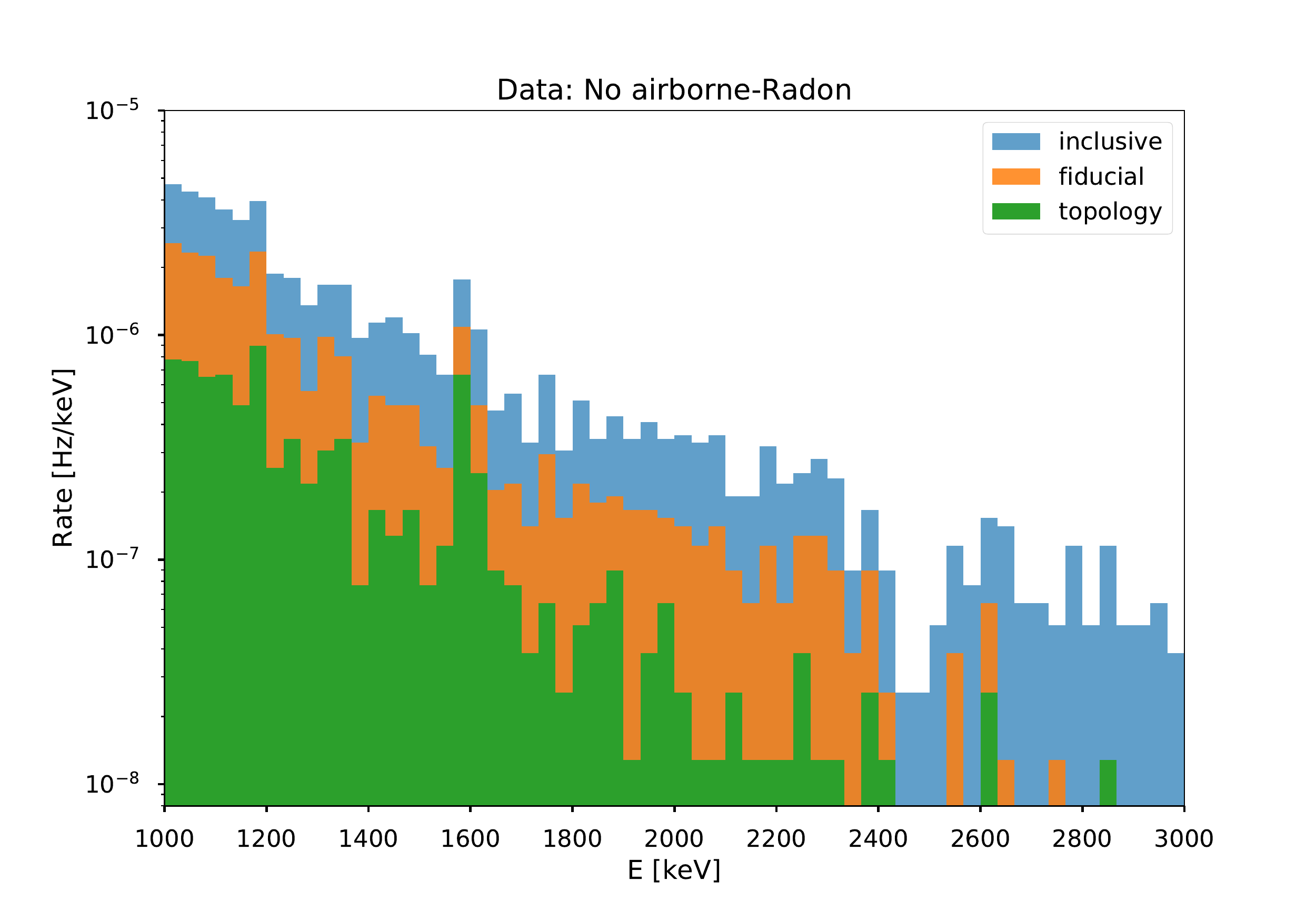}
\includegraphics[width=0.49\textwidth]{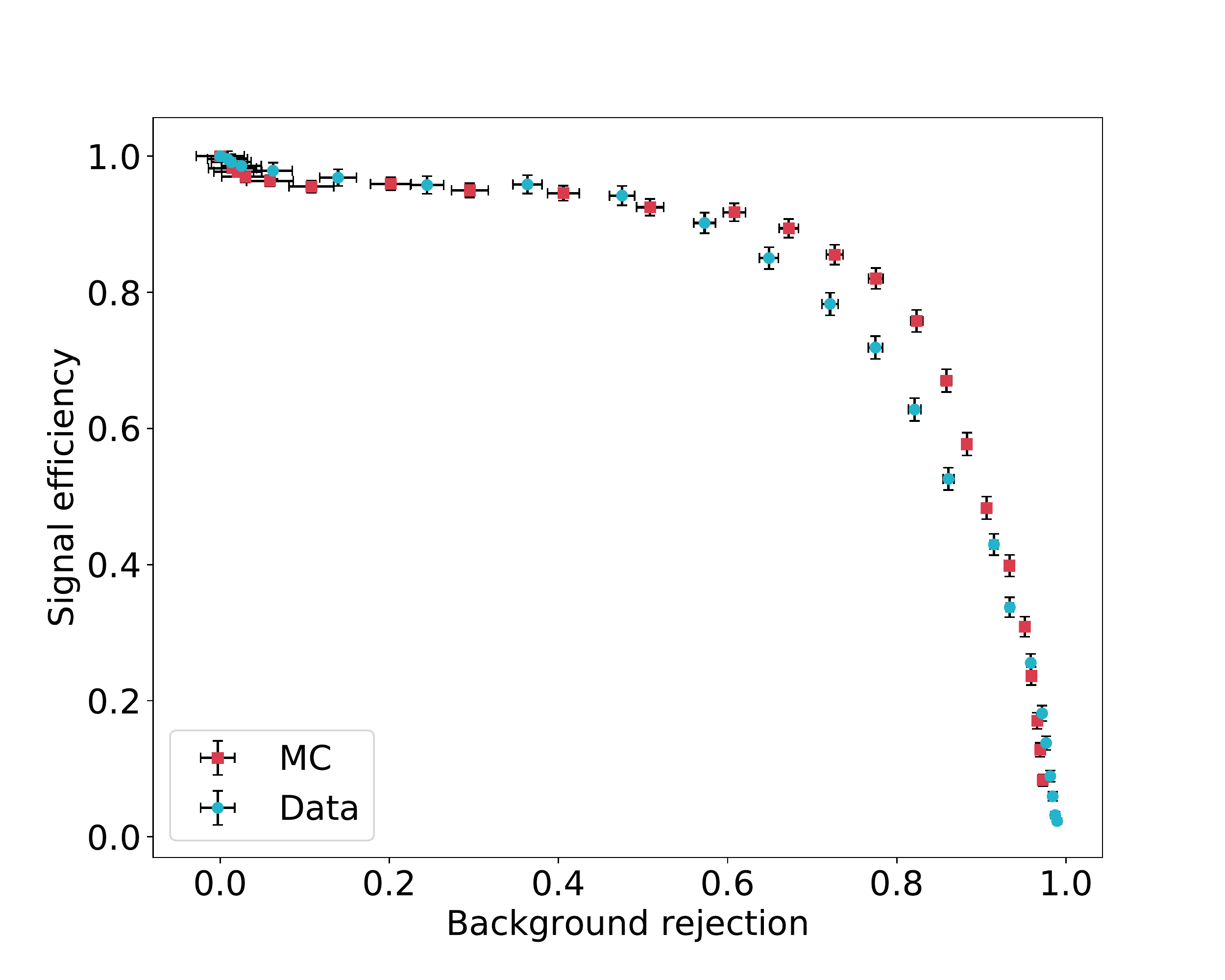}
\includegraphics[width=0.46\textwidth]{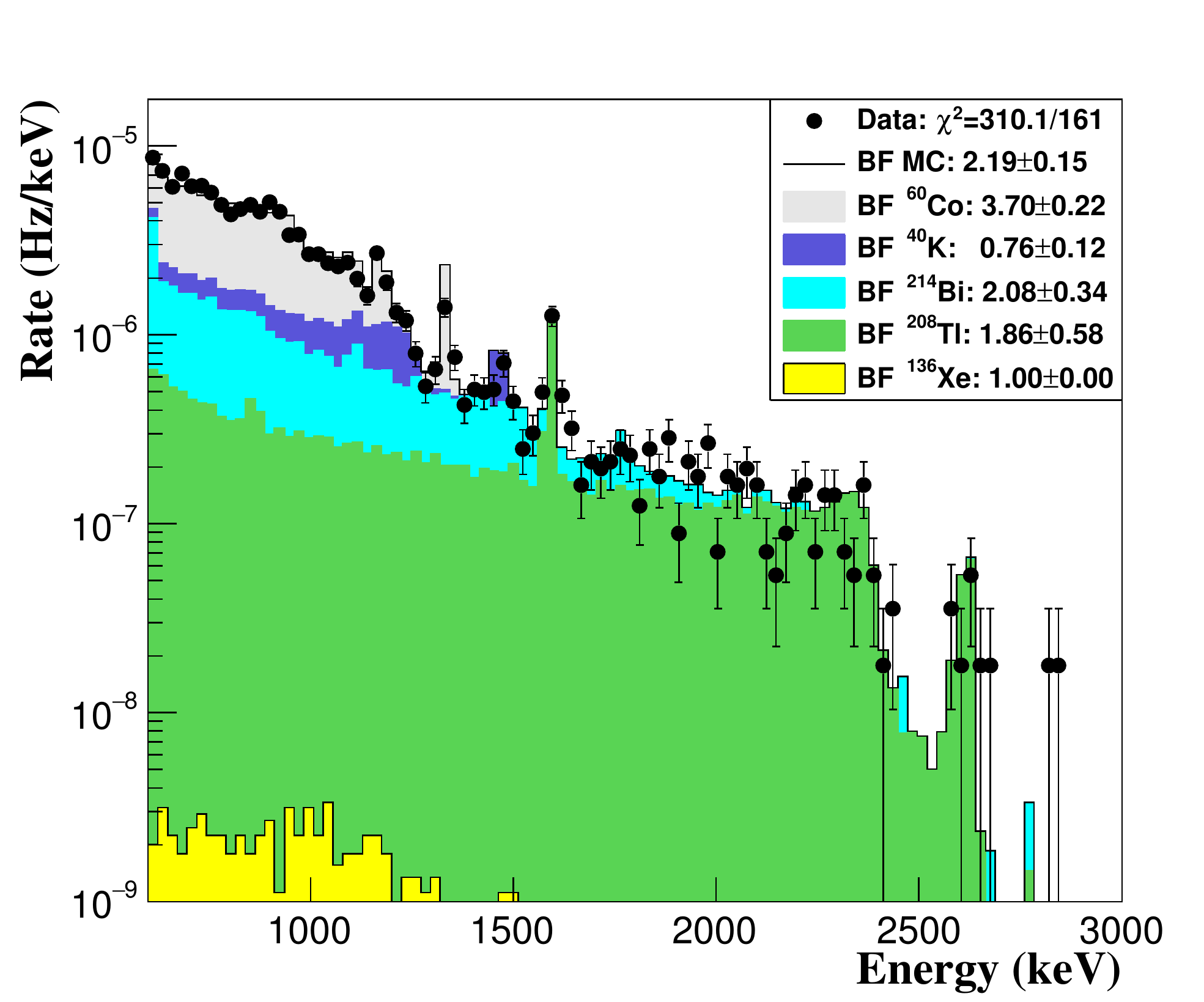}
\caption{A selection of the preliminary results obtained by \NEW. See text for details.}
\label{fig.newd}
\end{figure}

The detector operated with normal xenon during 7 months in 2017 at a pressure of \SI{7}{\bar} (Run II), and during 9 months in 2018 at a pressure of \SI{10}{\bar} (Run IV). Run V,  with enriched xenon will start in early 2019. Operation in Run II established a procedure to calibrate the detector with krypton decays ~\citep{Martinez-Lema:2018ibw}, and provided initial measurements of energy resolution,\citep{Renner:2018ttw}, electron drift parameters such as drift velocity, transversal and longitudinal diffusion \citep{Simon:2018vep} and a measurement of the impact of \RAD\ in the radioactive budget, that was found to be small \citep{Novella:2018ewv}. In addition,
the performance of the topological signal was measured from the data themselves. 
Figure \ref{fig.newd} shows a selection of preliminary results, including: resolution obtained at high energy fitting the \TL\ photopeak, which extrapolates to $0.85\%$ FWHM at \Qbb\ (top-left); rate (in \SI{}{Hz/keV}) of background events as a function of energy after 41.5~days of low background run, for three different event selections (top-right); signal efficiency versus background acceptance for the topological signature (bottom-left); fiducial background energy spectra, showing good agreement between data and Monte Carlo (bottom-right).

Run IV has demonstrated excellent operational stability, and a long electron lifetime, which in turn translates in improved resolution. 
Furthermore, the background rate of the detector has been reduced a factor of 4 with respect to Run II, thanks to operation in a radon-free atmosphere and enhanced external shielding. Preliminary results indicate good agreement between the measured data and the Monte Carlo background model, with an average rate of \SI{2}{\milli\hertz}. 

\subsection{The \NEXT\ detector}
\label{sec.next100}
%\begin{figure}[htb!]
%\centering
%\includegraphics[width=0.9\textwidth]{imgNEXT/next-100.png}
%\caption{\small The \NEXT\ detector.}
%\label{fig.next-100}
%\end{figure}

\begin{figure}[h!]
\centering
\includegraphics[width=1.0\textwidth]{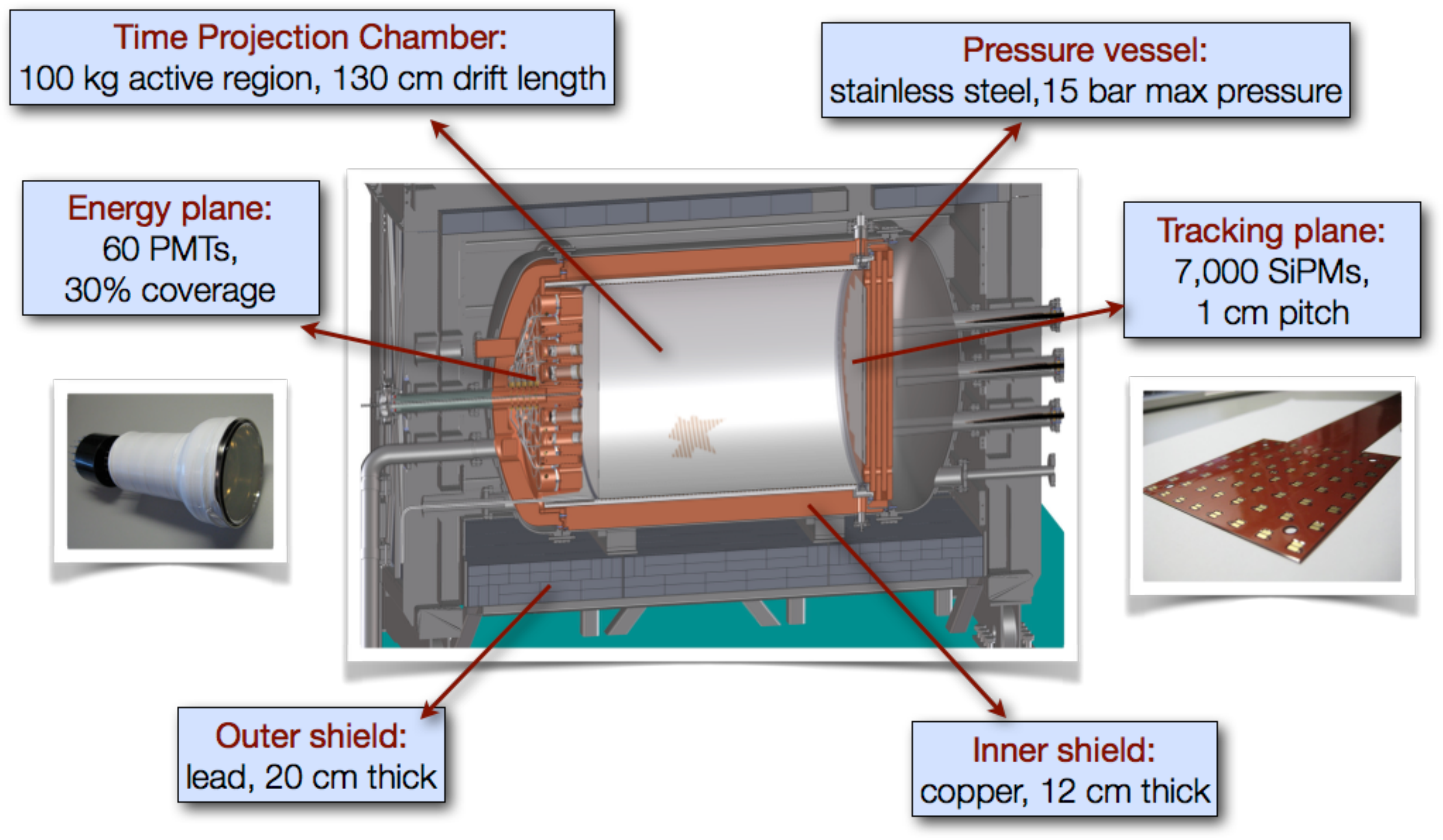}
\caption{\small The \NEXT\ detector.}
\label{fig.next-100}
\end{figure}

The \Next\ apparatus is shown schematically in figure \ref{fig.next-100}. The fiducial region is a cylinder of \NextTpcDiameter\ diameter and \NextTpcLength\ length (\NextFiducialVolume\ fiducial volume) holding a mass of \NextFiducialMass\ of xenon gas enriched at \XeEnrichment\ in \XE, and operating at \NextPressure.  The energy plane (EP) features \NextNumberOfPMT\ PMTs.  The tracking plane (TP) features an array of \NextNumberOfSiPM\ SiPMs.  \Next\ is essentially a 2:1 scale up version of \NEW.
%
%\begin{figure}[!htb]
%\centering
%\includegraphics[angle=0, width=0.7\textwidth]{img/BackgroundBudget.png}
%\caption{\Next\ background budget after selection.}
%\label{fig.nbb}
%\end{figure}

\begin{figure}[!h]
\centering
\includegraphics[angle=0, width=1.0\textwidth]{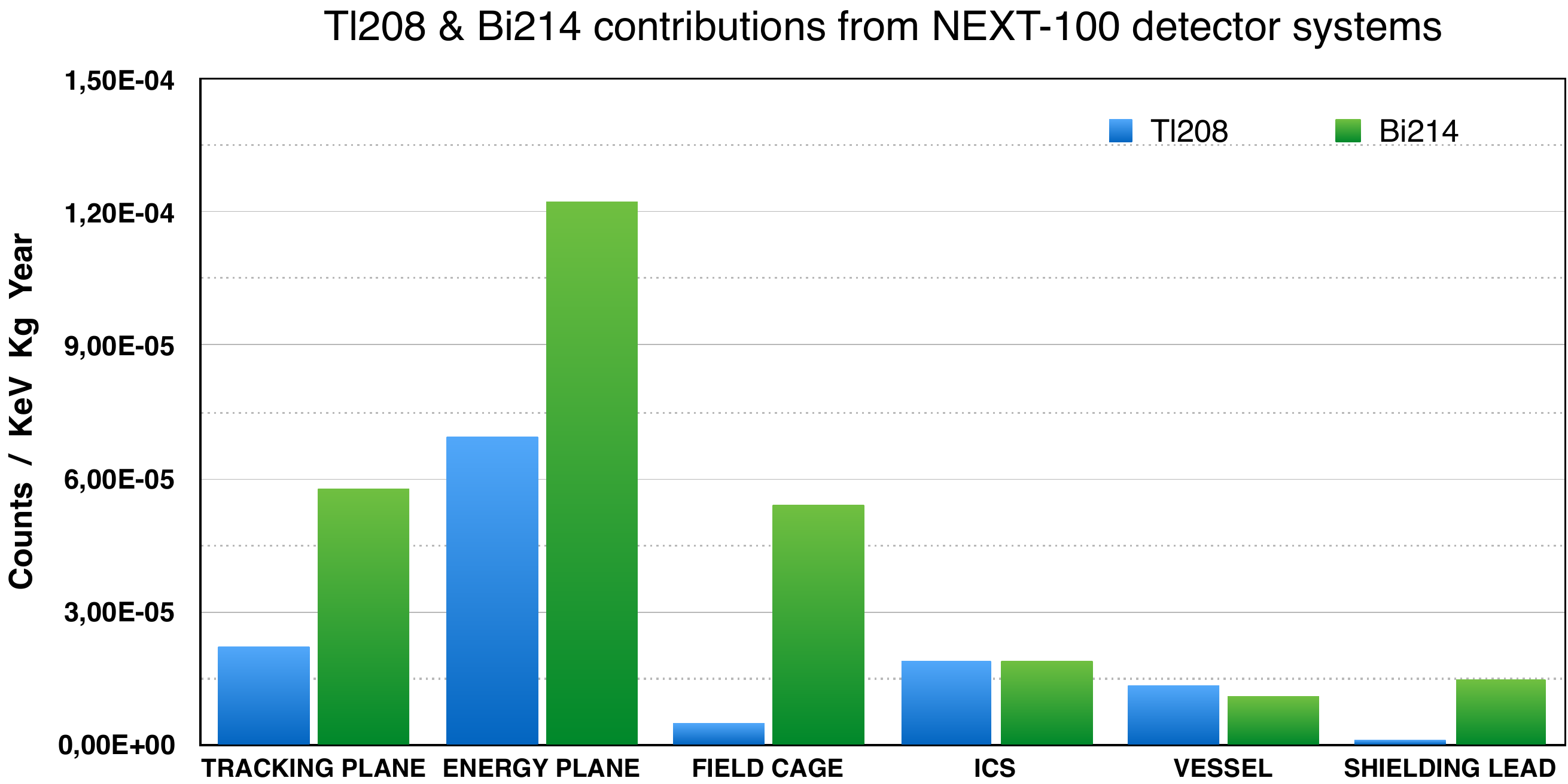}
\caption{\Next\ background budget after selection. Figure from \citep{javithesis}.}
\label{fig.nbb}
\end{figure}

The combination of excellent energy resolution and background rejection provided by the topological signature results in a very low background rate of \SI{4.5E-4}{counts\per\kg\per\keV\per\year} \citep{Martin-Albo:2015rhw}. The projected background in the ROI for \Next\ is \SI{0.7}{counts\per\year}, with the leading background sources being the PMTs and the substrates of the SiPMs as illustrated in \fig\ \ref{fig.nbb}. The overall efficiency of the detector is $32\%$. \Next\ is scheduled to start in operations in 2020. \Next\ will reach a sensitivity of \SI{1E+26}{\year} in \Tonu\ after a exposure of \SI{500}{\kg\year}. Although this is the same sensitivity achieved by KamLAND-Zen, the capability of NEXT-100 to provide a nearly background-free experiment at the \SI{100}{kg} scale, and the potential (discussed below) to improve its radioactive budget, resolution and topological signature so that background-free experiments at the ton scale are also possible, is the strongest asset of the experiment.

\subsection{Exploring the inverted hierarchy with \Ntk}
\label{sec.ntk}

%\begin{figure}[hb!]
%\centering
%\includegraphics[width=0.5\textwidth]{img/sensi-xenon.png}
%\caption{\small Sensitivity of a fully efficient \XE\ experiment as a function of the exposure, for different background rates.}
%\label{fig.Xe}
%\end{figure}

%\begin{figure}[hb!]
%\centering
%\includegraphics[width=0.45\textwidth]{img/voxelsTracks.pdf}
%\includegraphics[width=0.45\textwidth]{img/sigvsbg_DNN.pdf}
%\caption{\small Left: The three projections of a reconstructed Monte Carlo electron in \Next\ (top panels) and \Ntk\ (bottom panels). Right: Signal efficiency versus background rejection provided by the topological signature in both detectors. See text for details. Reproduced from \citep{Renner:2017ey}.}
%\label{fig.ts}
%\end{figure}

\begin{figure}[h!]
\centering
\includegraphics[width=1.0\textwidth]{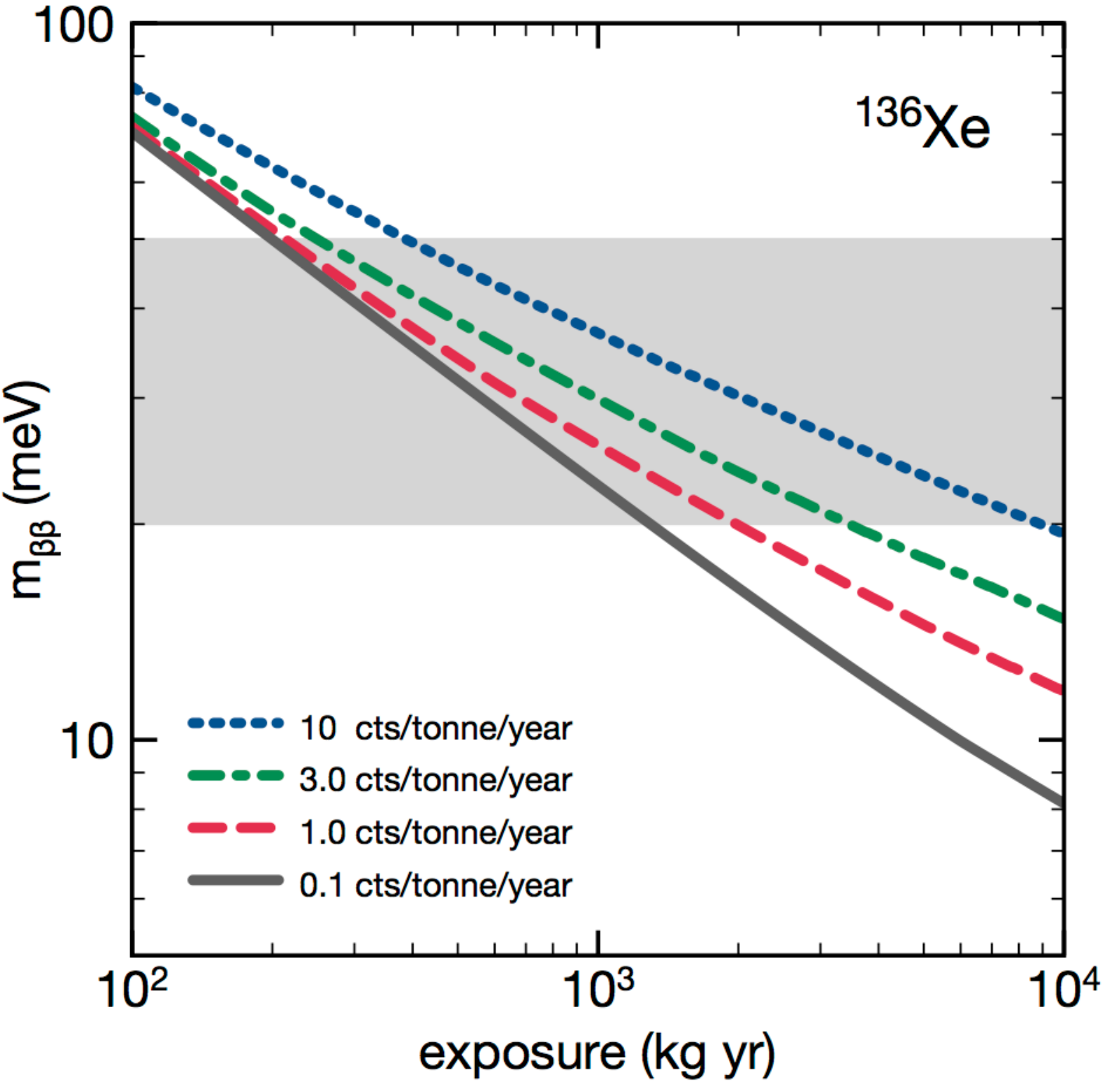}
\caption{\small Sensitivity of a fully efficient \XE\ experiment as a function of the exposure, for different background rates \citep{Martin-Albo:2015dza}.}
\label{fig.Xe}
\end{figure}

As discussed in the introduction, exploring the inverse hierarchy (IH) requires detectors with large masses (in the range of the ton) which should be ideally background free. 
The detrimental effects of the presence of background are illustrated in \fig\ ~\ref{fig.Xe}, which shows the sensitivity to \mbb\ (using a reasonable set of nuclear matrix elements) of a fully efficient \XE\ experiment as a function of exposure, for different background rates.  A fiducial exposure of almost \SI{2}{\tonne\yr} (corresponding to an actual exposure of \SI{6}{\tonne\yr}, when accounting for an efficiency of the order of $30\%$) is required for a full exploration of the inverse hierarchy for \BackgroundFreeLimit\ of background or less.  The exposure increases to \SI{3}{\tonne\yr} (\SI{9}{\tonne\yr}) for \AlmostBackgroundFreeLimit\ of background and degrades rapidly for larger backgrounds.  

The \HPXeEL\ technology can be scaled up to multi-tonne target masses while keeping extremely low levels of background by introducing several new technological advancements \citep{Cadenas:2017vcu}, including: a)  the replacement of PMTs (which are the leading source of background in \Next) with SiPMs, which are intrinsically radiopure, resistant to pressure and able to provide better light collection; b) operation of the detector at \NtkOperatingTemperature, not far from the gas triple point.  Operation in this regime has two marked advantages: i) it reduces the dark count rate of the SiPMs by a factor \num{300} and ii) it permits operation at a lower pressure  than \Next\ (\NextPressure\ at \NextTemperature) for the same gas density, as shown in \fig\ \ref{fig.isxe}.  Reducing the pressure, in turn, simplifies the construction of future larger detectors; c) operation of the detector with a low diffusion mixture, for example a 0.85/0.15 xenon/helium mixture ~\citep{Felkai:2017oeq}, which reduces the large transverse diffusion of natural xenon gas from \TransverseDiffusionPureXenon\ to \TransverseDiffusionXeHe, resulting in sharper reconstructed images for the electron trajectories and improving the performance of the topological signature~\citep{Renner:2017ey} ---other possible mixtures have been investigated in ~\citep{Henriques:2017rlj}---. 

\begin{figure}[h!]
\centering
\includegraphics[width=0.45\textwidth]{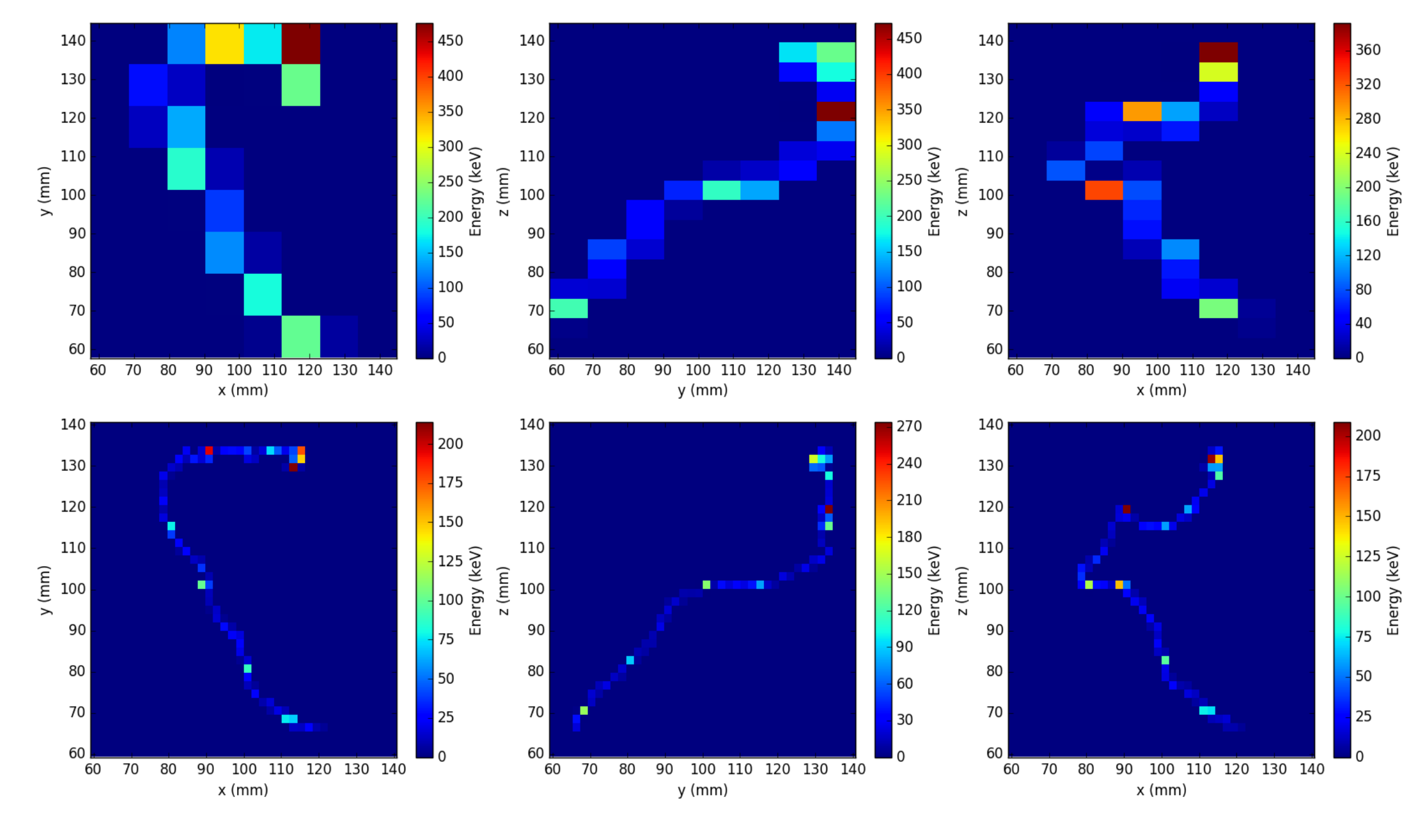}
\includegraphics[width=0.45\textwidth]{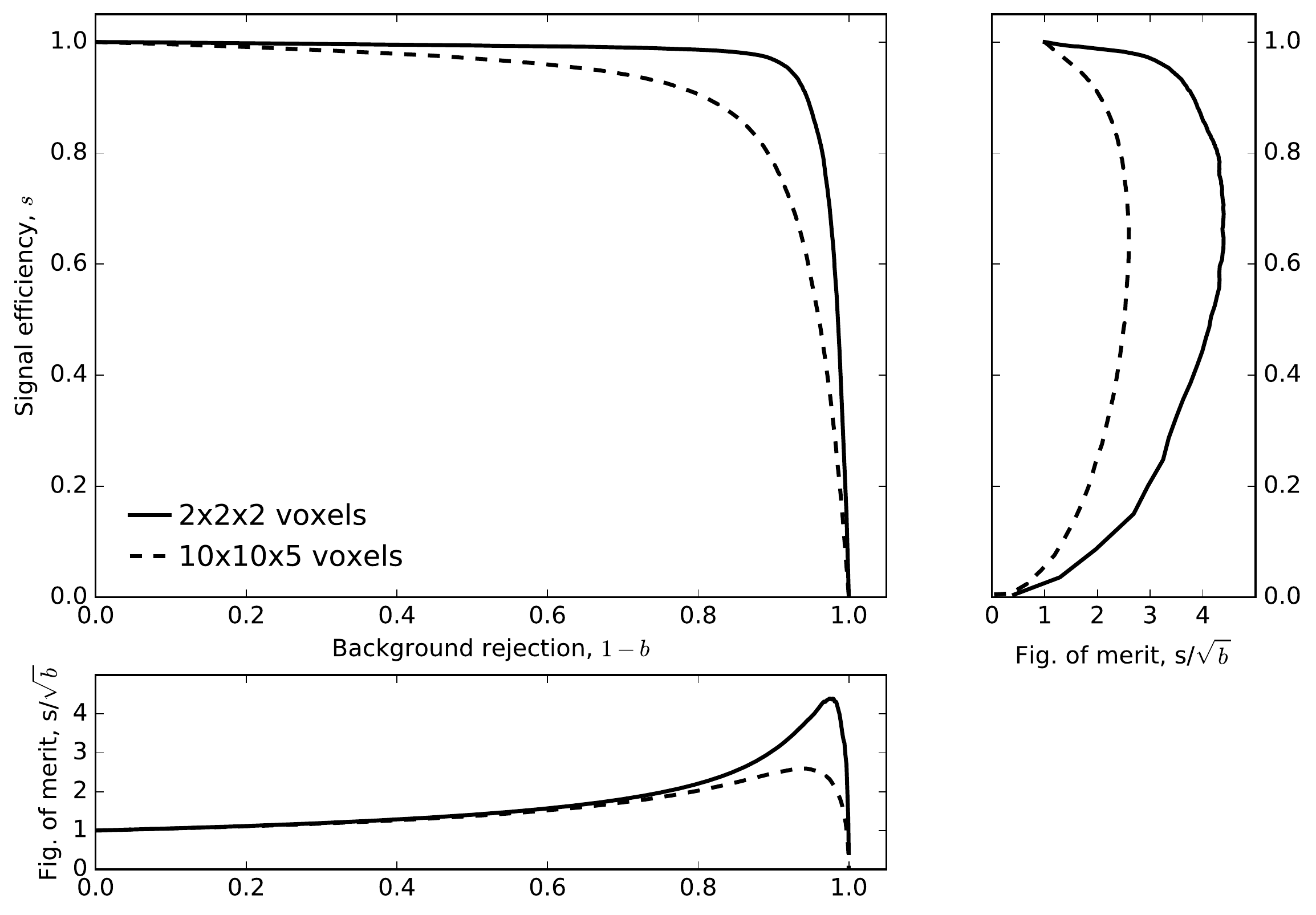}
\caption{\small Left: The three projections of a reconstructed Monte Carlo electron in \Next\ (top panels) and \Ntk\ (bottom panels). Right: Signal efficiency versus background rejection provided by the topological signature in both detectors. See text for details. Reproduced from \citep{Renner:2017ey}.}
\label{fig.ts}
\end{figure}

The improvement of the topological signature in low diffusion regime is illustrated in \fig\ \ref{fig.ts}. The left panel shows the three projections of a reconstructed Monte Carlo electron in  the case of high diffusion (top display) and low diffusion (bottom display). For the large transverse diffusion of natural xenon 
($\sim 10$ mm$/\sqrt{L}$, where $L$~is the length drifted by the electrons), the optimal size of the $x,y,z$~``voxels'', making up the reconstructed track is $10 \times 10 \times 5$~mm$^3$ ---for smaller voxels, the large diffusion results in disconnected tracks---. For the smaller transverse diffusion of a 0.85/0.15 xenon/helium mixture, ($\sim$\SI{2}{mm/\sqrt{L}}) the track can then be reconstructed with much smaller voxels, $2 \times 2 \times 2$~mm$^3$. The right panel of \fig\ \ref{fig.ts}  shows the expected efficiency versus rejection power of the topological cut separating single and double electrons for high diffusion (solid line, large voxels) and low diffusion (dashed line, small voxels). The figure of merit shown in figure maximizes in an efficiency close to $70\%$ in both case, but the background accepted is 
$6.6\%$ for high diffusion and $2.5\%$ for low diffusion. Thus, a reduction of 2.6 in background is achieved.  

The NEXT collaboration is planning a future detector, provisionally called \Ntk, that will incorporate all the above improvements, while deploying masses in the ton scale. If the tantalizing possibility of tagging the \Bapp\ ion is confirmed, \Ntk\ could incorporate a \Bapp\ tagging system. The target background of \Ntk\ is \SI{<1}{\ev\per\tonne\per\yr}, allowing the experiment to reach a sensitivity of \SI{1E27}{\yr} in \Tonu\ with an exposure of \SI{5.0}{\tonne\yr}.  This performance would improve the current state of the art by a factor of~\num{10}, opening the possibility for a discovery.  With \Bapp-tagging, \Ntk\ would be a truly background free experiment, capable to explore even further (or faster) the physical parameter space. 
Importantly, all the crucial technology improvements for \Ntk\ {\em can be demonstrated} by suitable upgrades of the \Next\ detector, which would permit testing each of the crucial steps leading to the ton-scale technology, namely cool gas operation, SiPM as energy sensors, low diffusion mixtures and barium tagging. 

\subsection{HXe TPCs based in electron multiplication: the NEXT-MM prototype}
\label{sbs.mm}

Although the baseline of the NEXT program is electroluminescence, a vigorous R\&D was conducted to assess the performance and potential of a HPXe TPC based in electron multiplication \citep{Gonzalez-Diaz:2015oba, Alvarez:2013kqa}. As extensively discussed in this report, the use of avalanche gain to amplify the ionization signal implies a cost in energy resolution with respect to electroluminescence. Such a cost could, in principle, be compensated by other factors. 

One of them is radiopurity. Micropattern structures such as micromegas are vey light and can be manufactured with radiopure components (\eg\ copper, kapton), thus offering a very radiopure readout system. Yet, the same applies to readouts based in SiPMs \citep{Alvarez:2014kvs}. Indeed, as discussed above, the main source of internal background in the \NEW\ and \Next\ detectors is the PMTs in the energy plane. However, getting rid of the PMTs is not easy, since they are needed to detect primary scintillation and thus provide a measurement of \tz. 
Indeed, the current trend in underground experiments both searching for Dark Matter and for \bbonu\ decays  is replacing the PMTs with SiPMs,  mounted on ultra-pure substrates. Such sensor arrangement will likely be as radiopure or more than micro-pattern-based readout planes. 

The use of micromegas alone does not provide a way to measure \tz\ and therefore one would need to revert to an asymmetric TPC, perhaps considering a detector that uses a plane of SiPMs behind the cathode to read primary scintillation and a micromegas-based readout located in the anode that would provide the event energy and topology. This solution has the same drawback already mentioned when discussing the early EXO proposal. One needs to build a large energy plane anyway, that could measure the energy with far better resolution than the solution being adopted. 

A second reason to consider a micro-pattern device readout is the possibility to combine a very fine pitch (\eg\ finely pixelized micro-bulk micromegas) with a low diffusion gas mixture, in order to improve the topological signature. In order to keep \tz\ such a low diffusion mixture also needs to shift the xenon VUV light to the visible or near UV spectrum, since the performance of micromegas in pure xenon deteriorates at high pressure, as discussed in
section \ref{sec.fundamentals}.

In \citep{Gonzalez-Diaz:2015oba, Alvarez:2013kqa}, the NEXT collaboration investigated the possibility that a mixture of xenon with Trimethylamine (TMA) could reduce electron diffusion while simultaneously displaying Penning effect and scintillation. Furthermore TMA could also provide charge neutralization of Ba$^{++}$~ to Ba$^+$, as needed for some Ba-tagging schemes (see section \ref{sec.smfi}). 

The studies were carried out with a medium-size HPXe TPC called NEXT-MM, of dimensions similar to NEXT-DEMO (\SI{73}{\liter}). 
The detector was instrumented with a large microbulk-Micromegas readout plane \citep{Alvarez:2013kqa}, covering an area of \SI{700}{\square\cm} and comprising $1152$ pixels of 8 $\times$ 8 \SI{}{mm^2}. 

The main results of those studies, using a mixture containing $2.2\%$ of TMA  and operating at \SI{10}{\bar} were:
\begin{enumerate}
\item The use of TMA allowed stable operation of the micromegas-based system at high pressure. 
\item TMA quenches the primary VUV xenon scintillation light even in very small concentrations. On the other hand, TMA scintillation was observed above \SI{250}{\nm} at the level of $100$ photons per MeV in \citep{Nakajima:2015cva}. In another study \citep{Trindade:2018}, TMA scintillation was not observed and a limit on re-emission w.r.t. the primary VUV light of $0.3\%$ was set. The conclusion is that in xenon-TMA mixtures it is not possible to use scintillation to measure \tz.
\item The use of TMA reduced by a large factor both the longitudinal and transverse diffusion, w.r.t. pure xenon. The measured values of $D_L^*$ at 1 bar 	~ranged between 340 
%$\mu \rm m  \sqrt{\rm bar}/\sqrt{\rm cm}$
$\mu \rm m /\sqrt{\rm cm}$
~and 649 
%$\mu \rm m  \sqrt{\rm bar}/\sqrt{\rm cm}$
$\mu \rm m  /\sqrt{\rm cm}$
~depending of the value of the reduced electric field, a reduction of 2-3 w.r.t. pure xenon. The measured value of the transverse diffusion was 
%$D_T^* \sim 250 \mu \rm m \sqrt{\rm bar}/\sqrt{\rm cm} $,
$D_T^* \sim 250 \mu \rm m /\sqrt{\rm cm} $,
 that is a factor 10 w.r.t. pure xenon.
\item The energy resolution (FWHM) ultimately achieved on the full fiducial volume was 14.6\% at \SI{30}{\keV}, with contributions from the limited S/N, sampling frequency and non-uniformity of the readout plane explaining the deterioration with respect to results obtained earlier in small setups ($9\%$). 
\item The calorimetric response to \SI{511}{\keV} and  \SI{1275}{\keV} electron tracks extrapolated to \Qbb\ gives energy resolutions varying from $3.2\%$ FWHM (for the best sector) to $3.9\%$ FWHM in the full TPC.
\item Due to the very low electron diffusion measured for the mixture (at the scale of 
\SI{1}{\mm} for \SI{1}{\m} drift), and the easiness at increasing the readout granularity, the technology offers the possibility of mm-accurate true-3D reconstruction of MeV-electron tracks on large detection volumes and at high pressure. Unfortunately, the lack of a source of double electrons in the system prevented from a full study characterizing the topological signature in this device as those carried out in NEXT-DEMO and \NEW. 
\end{enumerate}

In conclusion, this R\&D confirmed both the advantages ---excellent track reconstruction due to reduced diffusion--- and disadvantages ---worse energy resolution--- of using electron amplification, known since the pioneer work of the St. Gotthard TPC. The fact that TMA does not behave as a ``magical gas'' (its scintillation cannot be used to measure \tz), excludes its use for a future ton-scale experiment, unless alternative ways of measuring \tz\ are found. 

%On the other hand, the availability of mixtures capable to reduce xenon transverse diffusion without quenching the primary light ~\citep{Felkai:2017oeq} (the mixtures investigated in
 %~\citep{Henriques:2017rlj}, although partially quenching the primary light made still possible a \tz\ measurement), combined with the good performance of the topological signature in an HPXeEL and its excellent energy resolution make the use of 

\section{Other HPXe proposals}
\label{sec.axel}

\subsection{AXEL}

%\begin{figure}[hb!]
%\centering
%\includegraphics[width=0.6\textwidth]{img/elccc2.pdf}
%\caption{\small The ELCC concept.}
%\label{fig.elcc}
%\end{figure}

\begin{figure}[h!]
\centering
\includegraphics[width=1\textwidth]{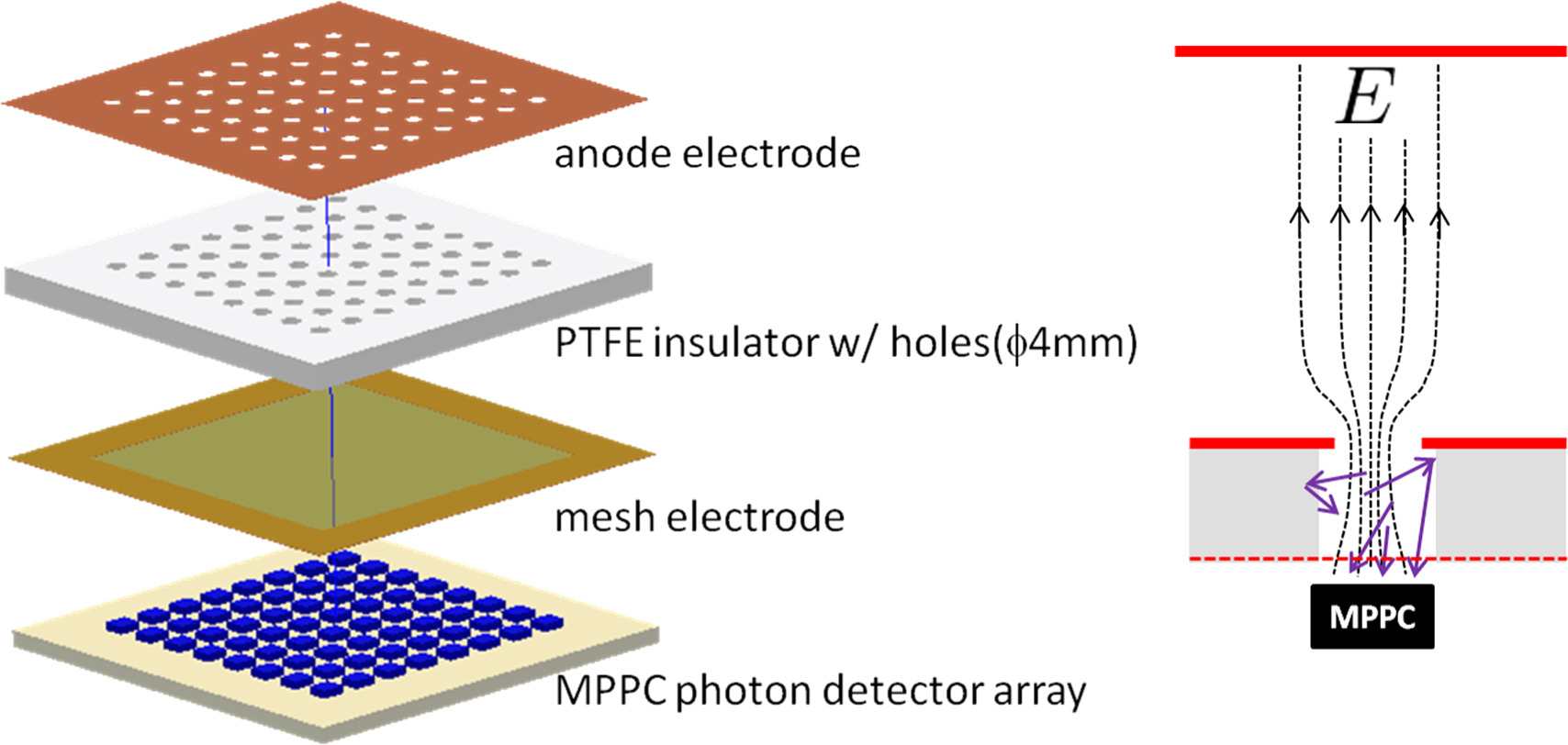}
\caption{\small The ELCC concept. Figure from \citep{Nakamura:2017tls}.}
\label{fig.elcc}
\end{figure}

AXEL is an R\&D lead by the U. of Kyoto, in Japan. The envisioned AXEL detector is almost identical to the NEXT design. Both are HPXe-EL TPCs, with an energy plane (AXEL assumes PMTs, as the \Next\ detector) and a tracking plane based in SiPMs. In the AXEL concept, however, the SiPMs are VUV sensitive and provide a measurement of the energy of the event in addition of a reconstruction of the topological signal. Rather than an open EL region, AXEL introduces the concept of Electroluminescence Light Collection Cell (ELCC), shown in \fig\ \ref{fig.elcc}. The ELCC consists of an anode plate, a supporting PTFE plate, a mesh and a plane of VUV SiPMs. The anode plate and PTFE have holes aligned with the SiPMs and define a cell structure. By applying high voltage between the anode plate and the mesh, ionized electrons are collected into the cells along the lines of electric field, and generate EL photons, which are detected by the SiPMs cell by cell. Because the EL region is contained in each cell, a measurement of the energy with this arrangement has a milder dependence on event position than in the case of the open grid used by NEXT. Notice, however, that the energy of the event in NEXT is measured with the PMT plane, while in AXEL the same PMT plane is only used to detect the primary scintillation light and thus measure \tz. This is due to the fact that the ELCC does not produce enough backward-going light, since only the photons moving along the narrow channel defined by the cell escape the structure, while all the others are absorbed. 

A small prototype has been developed by the AXEL collaboration. The detector is
\SI{6}{\cm} long and has a diameter of \SI{10}{\cm}.  The ELCC plane has 64 $3 \times 3$ Hamamatsu SiPMs
arranged in a $8 \times 8$~matrix and placed at \SI{7.7}{\mm} pitch. Two VUV sensitive PMTs, capable to operate up to \SI{10}{\bar} pressure provide \tz. The detector has operated at a pressure of  \SI{4}{\bar} and calibrated with low energy X-rays produced by a \COFiftySeven\ source. 

The energy resolution was measured using the xenon X-rays  \citep{Ban:2017ett}
(of energies \SI{29.8}{\keV}, \SI{33}{\keV}) as well as the photoelectric and escape peak from the \COFiftySeven\ source (energies of \SI{122}{\keV}, \SI{92}{\keV}).  %The fit to the four peaks was fitted to the simple statistical term $a \sqrt{E}$
The energy resolution as a function of the peak energy was fitted to the simple statistical model $a \sqrt{E}$ and also to a function of the form 
$a\sqrt{E + b E^2}$, which described better the data. The extrapolation to \Qbb\ was
$0.85\%$ using the $a \sqrt{E}$~law, and $2.03\%$ using the law with an additional constant term.  

Thus, in terms of energy resolution, the ELCC does not appear to bring an improvement over
the open grid used by NEXT (notice that the \NEW\ detector measures krypton X-rays, of 
energy \KrEnergy, with a resolution of $ 3.86 \pm 0.01$\% in the central region of the 
detector, which extrapolates to $0.5\%$ at \Qbb, to be compared with $4\%$ obtained by the AXEL prototype for the \SI{122}{\keV} gamma, which extrapolates to $0.9\%$ at \Qbb.). On the other hand PMTs are still needed to measure \tz\ and those PMTs could 
measure the energy with better resolution that AXEL has achieved so far. It remains to be 
seen if the ELCC brings an improvement to the topological signal. So far, the energies 
investigated by AXEL are too small to produce significant tracks.

On the other hand, the AXEL R\&D addresses two important points. One is the need to build 
very large EL structures for future ton-scale detectors. In that respect, the modular nature of 
the ELCC appears, a priori, well suited to scale up to large dimensions. The second point is 
the interest to measure the energy at the anode, if the cathode is to be used, in a future 
experiment for \Bapp\ tagging. However, the ELCC provides only a partial solution, since
AXEL still needs to instrument the cathode with PMTs to measure \tz. Further progress is to 
be expected with the planned larger prototype that the AXEL group plans to build in the near future. 
 
\subsection{PandaX-III}
\label{sec.pandax}

The PandaX-III collaboration \citep{Chen:2016qcd} has proposed the construction of a detector which is essentially a large-scale version of the NEXT-MM prototype discussed in section \ref{sec.next}, that is, a HPXe TPC based in electron amplification with a micromegas-based readout and a xenon-TMA mixture.  Their main argument for their technological choice are the reduction of background associated with the PMT energy plane, and the expected  enhancement of the topological signature. 

The performance of the topological signature expected in the detector has been quantified with Monte Carlo studies. A selection efficiency of $59\%$ for signal with a rejection of the background at the level of $97\%$ is found using a blob-search analysis. The results improve when using deep neural networks, which result in an $80\%$ signal efficiency and $98\%$ background rejection \citep{Han:2017fol}. This result is comparable with that obtained by NEXT using low diffusion mixtures ($70\%$ signal efficiency, $97.5\%$ background rejection). Indeed, both the PANDAX-III and the NEXT studies using DNNs show that the separation between single and double electrons in dense xenon gas reaches an intrinsic limit where about $2\%$ of the background events fake the signal event for perfect reconstruction. Those are events where multiple scattering or Bremsstrahlung in single, energetic electrons, ``fake'' an energetic blob at the beginning of the electron track. 

The parameters of the first PANDAX-III module are described in \citep{Han:2017fol}. The expected energy resolution (3\% at \Qbb) is consistent with the best results obtained by NEXT-MM. The efficiency ($35\%$) is typical of a HPXe TPC with track reconstruction and also consistent with that found by the NEXT detectors. 

%However, their expected background rate in the ROI ($10^{-4}$\ckky) is not consistent with the background expected by NEXT-100 ($4\times 10^{-4}$\ckky). While PANDA-X-III has no PMTs (thus reducing their radioactive budget by a factor 2 with respect to NEXT-100), and the performance of its topological rejection could improve that of NEXT-100 by another factor of 2 ---due to the reduced diffusion--- its energy resolution is at least 4 times worse than that of NEXT, implying, conservatively, a factor 4 increased background w.r.t. NEXT, given the proximity of the \BI\ and \TL\ peaks). On the other hand, the lack of \tz  implies that electrons emanating from \BI\ decays in the detector grids (in particular in the cathode) cannot be vetoed. In \citep{Novella:2018ewv} the effect of the Z cut in the background suppression is evaluated from the data themselves for the \NEW\ detector. The Z cut reduces the background by a factor 15. In the absence of Z cut a fraction of the background events is still eliminated by further cuts (\eg\ single track condition, double blob), but one still expects the rate of background in the ROI to increase by a factor 2--3 w.r.t. NEXT, and thus it is hard to understand how their overall background rate can be a factor 4 better. 

Comparing PANDAX-III with the SGTPC, we notice that both detectors are rather similar,  in particular concerning the performance of the topological signature, since both operate in the low diffusion regime (recall from section \ref{sec.history} the excellent performance of the topological signal of SGTPC). On the other hand, the energy resolution of the SGTPC was roughly a factor two worse than that projected by PANDAX-III, and none of the two detectors had \tz, thus one expects similar background level associated with radon degassing at the cathode. Yet, the SGTPC measured a background rate of $10^{-2}$\ckky, which is a factor 100 worse than the projected background level of PANDAX-III ($10^{-4}$\ckky). A naive expectation in terms of energy resolution would be a factor two improvement.  This suggests that the background rate estimation of PANDAX-III may be somewhat optimistic.

\section{Barium Tagging in a HPXe TPC}
\label{sec.smfi}
It has long been recognized that the detection of single barium ions emanating from the decay of \XE, when combined with a Gaussian energy resolution better than 2\% FWHM  ---needed to reject \bbtnu\ events which also produce barium ion---, could enable a background--free \bbonu\ experiment, since no conventional radioactive process can produce a barium ion in the gas xenon. 

A method to tag barium in a HPXe TPC was proposed in 2000 \citep{Danilov:2000pp}, 
following the idea pointed out in  \citep{Moe:1991ik} of using laser induced fluorescence to tag the presence of a
\Bap\ ion in xenon gas. 

%
%\begin{figure}[hb!]
%\centering
%\includegraphics[width=0.6\textwidth]{img/balevels.pdf}
%\caption{\small Atomic level scheme for \Bap\ ions.}
%\label{fig.bap}
%\end{figure}

The level structure of the \Bap\ ion shows a strong
\SI{493}{\nm} allowed transition, and therefore ground-state ions can be optically excited to the 
6${}^2P_{1/2}$ state from where they have substantial branching ratio (30\%) to decay into the metastable 5${}^4P_{3/2}$ state. A \Bap\ ion confined in a radio frequency trap can then be illuminated with suitable lasers to induce fluorescence. Specific \Bap\ detection is achieved by exciting the system back into the 6${}^2P_{1/2}$ state with 650 nm radiation and observing the blue photon from the decay to the ground state (70\% branching ratio). This transition has a spontaneous lifetime of \SI{8}{\ns} and radiates $6 \times 10^7$~photons/s.

While the technique has been established by atomic physicists since 1978
%W. Neuhauser, M. Hohenstatt, P. Toschek, and H. Dehmelt, Phys. Rev. Lett. 41 (1978) 233.
the application of it to a \bbonu\ experiment in a large HPXe TPC presents many formidable problem. The method proposed in \citep{Danilov:2000pp} assumed that the ion position in the TPC could be located {\em in flight} and illuminated with a 
pair of lasers tuned onto the appropriate frequencies and simultaneously steered to the place where the \bbonu\ candidate event was found. This is far from easy in a large HPXe TPC as those foreseen for the next-generation of neutrinoless double beta experiment. Additional complications were the detection of the fluorescence and the Doppler broadening of the transition line width at high pressure. Finally,  
barium resulting from double beta decay is initially highly ionized due to the disruptive departure of the two energetic electrons from the nucleus \cite{PhysRev.107.1646}. Rapid capture of electrons from neutral xenon is expected to reduce this charge state to \Bapp, which may then be further neutralized through electron-ion recombination.  Unlike in liquid xenon, where recombination is frequent and  the barium daughters are distributed across charge states~\cite{PhysRevC.92.045504}, recombination in the gas phase is minimal \cite{1997NIMPA.396..360B}, and thus Ba$^{++}$ is the expected outcome.
Thus, an additional transfer gas needs to be added to xenon, capable of transferring \Bapp\ 
to \Bap. TEA has been demonstrated to do the job \citep{Sinclair:2011zz} and similar gases such as TMA will probably also work. On the other hand, those very same gases quench the scintillation signal and thus the event cannot be located in the longitudinal coordinate. This fact alone makes the prospect for in-situ tagging very dim. 

Since 2000, the R\&D effort in barium-tagging has branched to two main lines. Tagging 
in liquid xenon \citep{Mong:2014iya} ---not discussed in this report---, and tagging in high-pressure gas with two main approaches: a) extracting the \Bapp\ ion to a secondary detection volume via funneling, and b) tagging the \Bapp\ ion in a suitable detector located in the cathode.

In fact, both approaches are not incompatible. The key notion proposed in \citep{Brunner:2014sfa} is to develop an RF ion-funnel to extract the \Bapp\ from the high pressure detector. If this is achieved the detection of the ion could proceed via guiding it to a quadrupole trap and using laser induced fluorescence or any other method. In particular, the ion could be guided to a \Bapp\ ion detector. On the other hand, the very same RF-carpets proposed for funneling the ion outside the HPXe can be used to guide it to a small region in the cathode itself, where they can be detected. Both approaches (in-situ versus extraction) have their cons and pros and both need still considerable R\&D (much of it synergic) to establish their feasibility. 

Yet, tagging ``in situ'', without lasers needs of a new detection technique. Such a technique, based in Single Molecule Fluorescence Imaging (SMFI) was proposed in \citep{nygrenbata}. 

SMFI is a technique invented by physicists and developed by biochemists that enables single-molecule sensitive, super-resolution microscopy. Among the applications of SMFI are the sensing of individual ions \cite{Lu2007}, demonstrated in various environments, including inside living cells \cite{stuurman2006imaging}.  
In SMFI, a thin layer containing potentially fluorescent molecules is repeatedly illuminated with a laser at frequencies in the blue or near ultraviolet range. The response of the molecule depends on whether they have captured a specific ion ---for example \Capp\ which is of great interest in neurological applications, given its role as neurotransmitter channel--- or not. Chelated molecules (molecules that have captured an ion) fluoresce strongly, while un-chelated (ion-free) molecules respond very weakly. Image-intensified CCD cameras are used to detect single photons and precisely identify and localize single molecules.

The NEXT collaboration is pursuing a program of R\&D to tag the barium ion using  SMFI techniques.  Since the \Bapp\ energy in high pressure gas is thermal, and charge exchange with xenon is highly energetically disfavored, the \Bapp\ state is expected to persist through drift to the anode plane.  For this reason, and because barium and calcium are congeners, dyes which have been developed for \Capp\ sensitivity for biochemistry applications provide a promising path toward barium tagging in HPXe.  In \citep{Jones:2016qiq} the properties of two such dyes, Fluo-3 and Fluo-4 were explored.  In the presence of  \Bapp\, excitation at \SI{488}{\nm} yielded strong emission peaking around \SI{525}{nm}, demonstrating the potential of these dyes to serve as barium tagging agents.  

%\begin{figure}[bth!]
%\begin{center}
%\label{Lego}
%\includegraphics[width=0.5\columnwidth]{img/LegoPlots-fixed.pdf} % Include the image placeholder.png
%\caption{A single Ba$^{++}$ candidate. A fixed region of the CCD camera is shown with 0.5~s exposure before (top) and after (bottom) photo-bleaching transition.\label{fig:legos}}
%\end{center}
%\end{figure}

A convincing proof-of-concept was carried out by the NEXT collaboration in
\citep{McDonald:2017izm}. The experiment managed to resolve individual 
\Bapp\ ions on a scanning surface using an SMFI-based sensor.  

The SMFI sensor concept uses a thin quartz plate with surface-bound fluorescent indicators, continuously illuminated with excitation light and monitored by an EM-CCD camera.  It is anticipated that such a sensor would form the basis for a \Bapp\ detection system in HPXe, with ions delivered to a few $\sim$\SI{1}{mm}$^2$ sensing surfaces, first via drift to the cathode and then transversely by RF-carpet \citep{ARAI201456}, a method already demonstrated at large scales \citep{Gehring2016221} and for barium transport in HPXe \citep{Brunner:2014sfa}, as discussed above.

To demonstrate single \Bapp\ the proof-of-concept imaged individual near-surface \Bapp\ ions from dilute barium salt solutions using total internal reflection fluorescence (TIRF) microscopy \citep{Burghardt2012}. The fluorophores used as detectors were fixed at the sensor surface.  This emulates the conditions in a HPXe TPC detector, where the ions will drift to the sensor plate and adhere to fluorophores immobilized there.  

\begin{figure}[h!]
\begin{center}
\includegraphics[width=1.0\textwidth]{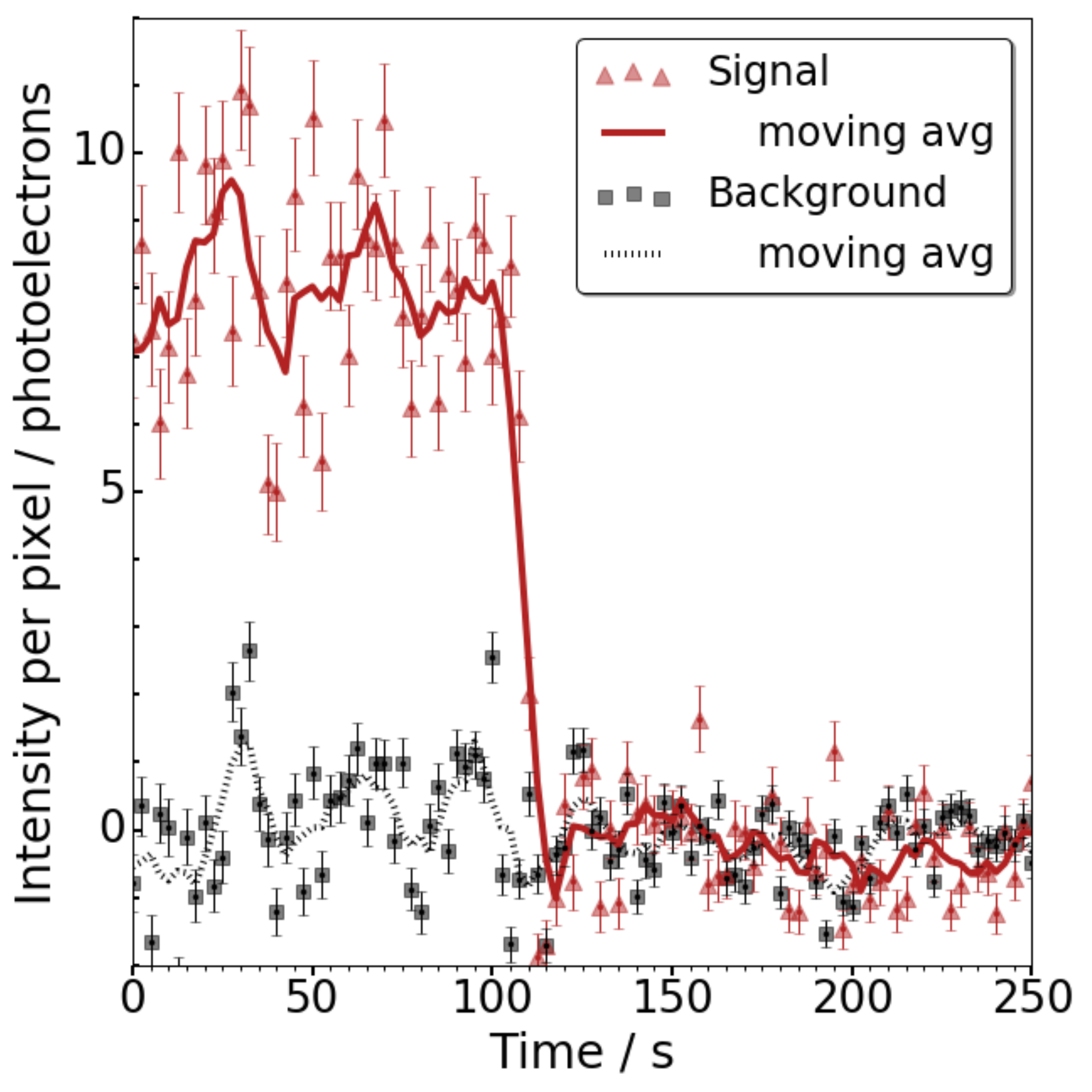} 
\caption{Fluorescence trajectory for one candidate in a barium-spiked sample. ``Signal'' shows the average activity in 5$\times$5 pixels centered on the local maximum. ``Background'' shows the average in the 56 surrounding. The single step photo-bleach is characteristic of single molecule fluorescence. Figure from \citep{McDonald:2017izm}. \label{fig:photobleach}}
\end{center}
\end{figure}

The hallmark of single molecule fluorescence is a sudden discrete photo-bleaching transition \cite{Habuchi05072005}.  This occurs when the fluorophore transitions from a fluorescent to a non-fluorescent state, usually via interaction with reactive oxygen species  \cite{thomas2000comparison}.  This discrete transition signifies the presence of a single fluor, rather than a site with multiple fluors contributing. The \SI{375}{s} scan time is significantly longer than the typical photo-bleaching time of Fluo-3 at this laser power \cite{thomas2000comparison}, so the majority of spots are observed to bleach in our samples. A typical near-surface fluorescence trajectory is shown in figure \ref{fig:photobleach}.  %One 0.5~s exposure of this spot directly before the step and one 0.5~s exposure directly after the step are shown in Fig.~\ref{fig:legos}.
In summary, the NEXT proof of concept shows that SMFI can be used to resolve individual \Bapp\ ions at surfaces via TIRF microscopy.  \Bapp\ ions have been detected above a background of free residual ions at $12.9 \sigma$ statistical significance, with individual ions spatially resolved and observed to exhibit single-step photo-bleaching trajectories characteristic of single molecules.

An SMFI sensor in a future HPXe TPC will differ from the apparatus described here in a few key ways.  First, the fluorophores will be surface-tethered, and not embedded in a thick sample. Thus, only near-surface bright spots are expected, and offline separation from the deeper background fluors will not be necessary. Second, the target signature will be appearance of a new candidate over a pre-characterized background, coincident in a spatio-temporal region with an \bbonu\ candidate in the TPC. In this case, only the ability to resolve appearance of a new ion is important, and the spatial localization of individual ion candidates demonstrated here shows that many can be recorded on the same sensor before saturation.  Third, the micro-environment around the Fluor will be different, being immobilized on a dry surface rather than within a PVA matrix, and this may modify chelation and fluorescence properties of the Fluor.  Finally, the extent to which photo-bleaching will be active in a clean HPXe environment is unknown.  
Thus, there is still a long and uncertain R\&D road ahead before barium tagging can be successfully implemented in a HPXe detector searching for \bbonu\ decays.

\section{Outlook}
\label{sec.conclu}

The Time Projection Chamber (TPC), invented by D. Nygren in 1975 revolutionized the imaging of charged particles in gaseous detectors.  This article has presented a review of the application of high pressure xenon (HPXe) TPCs to the search for neutrinoless double beta decay (\bbonu) processes. This field has been active for more than half a century, but its importance has been re-asserted with the discovery of neutrino oscillations and the implications that neutrinos have mass. Massive neutrinos could be Majorana particles, identical to their own antiparticles, a fact that would be unambiguously proven if \bbonu\ processes are observed. 

The simplest mechanism to mediate \bbonu\ decays (the exchange of a light neutrino) provides a rationale to evaluate the state-of-the-art and prospects of the field. The current generation of \bbonu\ experiments have explored half-lives in the vicinity of 
\SI{10E+26}{\yr}, corresponding to effective neutrino masses ---in the case of xenon and for the most favorable nuclear matrix element--- in the vicinity of \SI{60}{\meV}. The next generation of experiments aims to reach half-lives of at least \SI{10E+27}{\yr} and thus explore effective neutrino masses of up to  \SI{20}{\meV}, with a significant probability of making a discovery.

To be sensitive to \bbonu\ half-lives in the range of \SI{10E+27}{\yr} and more, it is mandatory that the next-generation experiments deploy exposures in the range of up to 
\SI{10}{ton\cdot\yr}, and a background rate in the ROI as close to zero as possible. This tremendous challenge all but forces a paradigm shift in the experimental techniques of the field.

High Pressure Xenon TPCs offer one of the most attractive approaches to address such a daunting challenge thanks to the combination of: a) excellent energy resolution; b) the availability of a robust topological signature that allows full fiducialization of the events, identification of satellite clusters (signaling Compton scatters and other background processes) and the capability to separate single (background) from double (signal) electrons. HPXe TPCs can be built with ultra-pure materials and are scalable to large masses. On top of that, the possibility to tag the daughter \Bapp\ ion produced in the decay 
$ \textrm{Xe} \rightarrow \textrm{Ba}^{++} + 2 \textrm{e}^- (+ 2\nu)$, may provide a way to build truly background-free experiments.

The use of xenon for \bbonu\ searches was proposed as early as 1961, and the realization that electroluminescence (EL) was a promising way to achieve excellent energy resolution dates from 1975 (as part of the first detector proposal, a bubble chamber), and was revisited in 1983. However, the first generation of HPXe detectors (the Baksan and Milano experiment) did not use EL but electron amplification. The first true HPXe time projection chamber ---the St. Gotthard TPC--- was also based in avalanche gain, and used a mixture 0.95-0.05 Xe/CH$_4$~to stabilize the gas, which quenched the primary scintillation (thus the detector had to make do without \tz) and resulted in mediocre energy resolution ($\sim$ $7\%$ at \Qbb). However, the SGTPC also proved the robust topological signal available to the technology and boasted the lowest background rate level ($10^{-2}\ckky$) of its time.

And yet, after the SGTPC the technology was frozen for two decades, until it was resurrected again by the NEXT experiment, which incorporated D. Nygren's proposal to build an EL TPC, along with technological solutions developed by J. White and one of the authors of this review (JJGC). The NEXT program has built two large prototypes and a detector of the size of the SGTPC (\NEW), which is currently taking data at the Canfranc Underground Laboratory and has already shown excellent energy resolution ($1\%$ FWHM at \Qbb), and a robust topological signal, along with a well understood background model. The \Next\ detector, currently in early phase of construction (scheduled to start data taking in 2020), can achieve a sensitivity competitive with the best experiments of the current generation while at the same time demonstrating the potential of the technology for a background free experiment. 

Two other projects are proposing HPXe TPCs for \bbonu\ searches. One is AXEL, in Japan, which has so far developed a small prototype in which the interesting concept of Electroluminescence Light Collection Cell (ELCC) is explored. The other is PANDAX-III, which proposes what is essentially a large-scale version of the SGTPC, using micromegas (rather than wires and pads) for readout, and TMA (rather than CH$_4$) as quencher. As it was the case with SGTPC, PANDAX-III is expected to feature an excellent topological signature but has the double handicap of a mediocre resolution (3--4 \% FWHM according to the R\&D measurements done in the context of the NEXT collaboration) and lack of \tz. 

This review has also summarized the state-of-the-art of \Bapp\ tagging in gaseous HPXe detector. R\&D has been conducted in this front by the EXO collaboration (which is also studying \Bapp\ tagging in liquid and solid xenon) and by NEXT. The EXO-gas collaboration has studied the possibility to extract the \Bapp\ ion out of the main chamber through funneling, while NEXT is studying the the potential of applying Single Molecular Fluorescence Imaging, SMFI (also proposed by Dave Nygren) to barium tagging with very promising initial results. 

In summary it appears  that the HPXe technology, in particular using electroluminescence and barium tagging, can assert itself as one of the major possibilities for the next generation of \bbonu\ experiments.

\section*{Conflict of Interest Statement}
%All financial, commercial or other relationships that might be perceived by the academic community as representing a potential conflict of interest must be disclosed. If no such relationship exists, authors will be asked to confirm the following statement: 

The authors declare that the research was conducted in the absence of any commercial or financial relationships that could be construed as a potential conflict of interest.

%\section*{Author Contributions}
%
%The Author Contributions section is mandatory for all articles, including articles by sole authors. If an appropriate statement is not provided on submission, a standard one will be inserted during the production process. The Author Contributions statement must describe the contributions of individual authors referred to by their initials and, in doing so, all authors agree to be accountable for the content of the work. Please see  \href{http://home.frontiersin.org/about/author-guidelines#AuthorandContributors}{here} for full authorship criteria.

\section*{Funding}
This research work has been supported by the European Research Council (ERC) under the Advanced Grant 339787-NEXT the Ministerio de Econom\'ia y Competitividad of Spain under grants FIS2014-53371-C04, and the GVA of Spain under grants PROMETEO/2016/120 and SEJI/2017/011. 

\section*{Acknowledgments}

We gratefully acknowledge the privilege of working and learning from Dave Nygren, and the late James White. We would also like to acknowledge Jos\'e \'Angel Hernando, Michel Sorel, Pau Novella, Joaquim dos Santos, Lior Arazi and our colleagues of the NEXT collaboration for many fruitful discussions. We also acknowledge Oliviero Cremonesi and Alexander Barabash for their kind input on the historical development of the HPXe technology. 

%Diego Gonz\'alez-Diaz for  illuminating discussions. 

%\section*{Supplemental Data}
% \href{http://home.frontiersin.org/about/author-guidelines#SupplementaryMaterial}{Supplementary Material} should be uploaded separately on submission, if there are Supplementary Figures, please include the caption in the same file as the figure. LaTeX Supplementary Material templates can be found in the Frontiers LaTeX folder.
%
%\section*{Data Availability Statement}
%The datasets [GENERATED/ANALYZED] for this study can be found in the [NAME OF REPOSITORY] [LINK].
% Please see the availability of data guidelines for more information, at https://www.frontiersin.org/about/author-guidelines#AvailabilityofData

\bibliographystyle{frontiersinSCNS_ENG_HUMS} % for Science, Engineering and Humanities and Social Sciences articles, for Humanities and Social Sciences articles please include page numbers in the in-text citations
\bibliography{hpxenf}

%%% Make sure to upload the bib file along with the tex file and PDF
%%% Please see the test.bib file for some examples of references

\section*{Figure captions}

%%% Please be aware that for original research articles we only permit a combined number of 15 figures and tables, one figure with multiple subfigures will count as only one figure.
%%% Use this if adding the figures directly in the mansucript, if so, please remember to also upload the files when submitting your article
%%% There is no need for adding the file termination, as long as you indicate where the file is saved. In the examples below the files (logo1.eps and logos.eps) are in the Frontiers LaTeX folder
%%% If using *.tif files convert them to .jpg or .png
%%%  NB logo1.eps is required in the path in order to correctly compile front page header %%%

%%%

%\begin{figure}[h!]
%\begin{center}
%\includegraphics[width=1.0\textwidth]{img2/GSC.pdf}
%\includegraphics[width=0.8\textwidth]{img/PC.pdf} 
%\end{center}
%\caption{\small Principle of a Gas Proportional Scintillation Counter.}
%\label{fig:GPSC} 
%\end{figure}

%\begin{figure}[h!]
%\centering
%\includegraphics[width=1.0\textwidth]{img/MicromegasResolution.pdf} 
%\caption{\small Resolution of a micro-bulk Micromegas as a function
%of the pressure for 22.1 keV photons. The resolution varies between 12\% at 1 bar ($\sim$ ) about 1\% at \Qbb, to
% 32\% at 10 bar, 3\% at \Qbb. The results were measured as a part of the NEXT collaboration R\&D
% \cite{Balan:2010kx}. }
%\label{fig.mm}			
%\end{figure}

%\begin{figure}[h!]
%\centering
%\includegraphics[width=0.8\textwidth]{img2/blob1_blob2.pdf}
%\caption{\small The bottom panel shows the energy of the blob of less energy versus the energy of the blob of high energy for a background electron propagating in an HPXe ---perfect track reconstruction---, while the top plot shows the same plot of the 2 electrons produced in a \bbonu\ propagating under the same conditions. In the first case the energy of the lower energy blob is much smaller than the energy of the higher energy blob, in the second case both are roughly the same.}
%\label{fig:bib2}
%\end{figure}

%\begin{figure}[h!]
%\centering
%\includegraphics[width=1.0\textwidth]{img2/geom_rejection.pdf}
%\caption{\small Charged particle backgrounds entering the detector active volume
%		can be rejected with complete 3D-reconstruction (top left). The mean
%		free path of xenon for the high-energy gammas emitted in \Bi\ and \Tl\
%		decays is $>3$ m, and thus many of them cross the detector without interacting
%		(top right). Also, \Bi\ and \Tl\ decay products include low-energy gammas
%		which interact in the vetoed region close to the chamber walls (bottom left).
%		Only those background events with tracks fully-contained within the fiducial volume
%		may mimic the signal (bottom right).
%\label{fig:geom_rejection}}
%\end{figure}

%\begin{figure}[h!]
%\begin{center}
%\includegraphics[width=1.0\textwidth]{img2/tracksSG.pdf}
%\end{center}
%\caption{\small A typical ``two electron'' candidate found in the data of the SGTPC: the XZ and YZ projection, as well as the extracted X-Y projections (in the lower frame) are drawn. Scales are in cm. The time evolution of the anode signal is displayed on the right. The $\beta\beta$-candidate exhibits ``blobs'' at both ends of a continuous track. (from \citep{Luscher:1998sd}).}
%\label{fig:tracksSG} 
%\end{figure}

\end{document}